\documentclass[preprint,aps,superscriptaddress,showpacs,nofootinbib]{revtex4-1}

\usepackage{amscd}
\usepackage{amsmath,amssymb,amsthm,mathrsfs,amsfonts,dsfont}
\usepackage{subfigure, epsfig}
\usepackage{braket}
\usepackage{bm}
\usepackage{enumerate}

\usepackage{graphicx}   
\usepackage{epsfig}
\usepackage{subfigure}
\usepackage{color}
\usepackage[10pt]{moresize}
\usepackage{longtable}

\newcommand\oprod[2]{\ensuremath{|#1\rangle\langle#2|}}
\newcommand\mean[1]{\ensuremath{\langle #1 \rangle}}

\begin{document}

\title{Experimental Twin-Field Quantum Key Distribution Through Sending-or-Not-Sending}

\author{Yang Liu}
\affiliation{Shanghai Branch, National Laboratory for Physical Sciences at Microscale and Department of Modern Physics, University of Science and Technology of China, Shanghai 201315, P.~R.~China}
\affiliation{Shanghai Branch, CAS Center for Excellence and Synergetic Innovation Center in Quantum Information and Quantum Physics, University of Science and Technology of China, Shanghai 201315, P.~R.~China}
\affiliation{Jinan Institute of Quantum Technology, Jinan, Shandong 250101, P.~R.~China}

\author{Zong-Wen Yu}
\affiliation{State Key Laboratory of Low Dimensional Quantum Physics, Department of Physics, Tsinghua University, Beijing 100084, P.~R.~China}
\affiliation{Data Communication Science and Technology Research Institute, Beijing 100191, P.~R.~China}

\author{Weijun Zhang}
\affiliation{State Key Laboratory of Functional Materials for Informatics, Shanghai Institute of Microsystem and Information Technology, Chinese Academy of Sciences, Shanghai 200050, P.~R.~China}

\author{Jian-Yu Guan}
\author{Jiu-Peng Chen}
\author{Chi Zhang}
\affiliation{Shanghai Branch, National Laboratory for Physical Sciences at Microscale and Department of Modern Physics, University of Science and Technology of China, Shanghai 201315, P.~R.~China}
\affiliation{Shanghai Branch, CAS Center for Excellence and Synergetic Innovation Center in Quantum Information and Quantum Physics, University of Science and Technology of China, Shanghai 201315, P.~R.~China}

\author{Xiao-Long Hu}
\affiliation{State Key Laboratory of Low Dimensional Quantum Physics, Department of Physics, Tsinghua University, Beijing 100084, P.~R.~China}

\author{Hao Li}
\affiliation{State Key Laboratory of Functional Materials for Informatics, Shanghai Institute of Microsystem and Information Technology, Chinese Academy of Sciences, Shanghai 200050, P.~R.~China}

\author{Cong Jiang}
\affiliation{State Key Laboratory of Low Dimensional Quantum Physics, Department of Physics, Tsinghua University, Beijing 100084, P.~R.~China}

\author{Jin Lin}
\author{Teng-Yun Chen}
\affiliation{Shanghai Branch, National Laboratory for Physical Sciences at Microscale and Department of Modern Physics, University of Science and Technology of China, Shanghai 201315, P.~R.~China}
\affiliation{Shanghai Branch, CAS Center for Excellence and Synergetic Innovation Center in Quantum Information and Quantum Physics, University of Science and Technology of China, Shanghai 201315, P.~R.~China}

\author{Lixing You}
\author{Zhen Wang}
\affiliation{State Key Laboratory of Functional Materials for Informatics, Shanghai Institute of Microsystem and Information Technology, Chinese Academy of Sciences, Shanghai 200050, P.~R.~China}

\author{Xiang-Bin Wang}
\affiliation{Shanghai Branch, CAS Center for Excellence and Synergetic Innovation Center in Quantum Information and Quantum Physics, University of Science and Technology of China, Shanghai 201315, P.~R.~China}
\affiliation{Jinan Institute of Quantum Technology, Jinan, Shandong 250101, P.~R.~China}
\affiliation{State Key Laboratory of Low Dimensional Quantum Physics, Department of Physics, Tsinghua University, Beijing 100084, P.~R.~China}

\author{Qiang Zhang}
\author{Jian-Wei Pan}
\affiliation{Shanghai Branch, National Laboratory for Physical Sciences at Microscale and Department of Modern Physics, University of Science and Technology of China, Shanghai 201315, P.~R.~China}
\affiliation{Shanghai Branch, CAS Center for Excellence and Synergetic Innovation Center in Quantum Information and Quantum Physics, University of Science and Technology of China, Shanghai 201315, P.~R.~China}

\begin{abstract}
Channel loss seems to be the most severe limitation on the practical application of long distance quantum key distribution. The idea of twin-field quantum key distribution can improve the key rate from the linear scale of channel loss in the traditional decoy-state method to the square root scale of the channel transmittance. However, the technical demanding is rather tough because it requests single photon level interference of two remote independent lasers. Here, we adopt the technology developed in the frequency and time transfer to lock two independent lasers' wavelengths and utilize additional phase reference light to estimate and compensate the fiber fluctuation. Further with a single photon detector with high detection rate, we demonstrate twin field quantum key distribution through the sending-or-not-sending protocol with realistic phase drift over 300 km optical fiber spools. We calculate the secure key rates with finite size effect. The secure key rate at 300 km ($1.96\times10^{-6}$) is higher than that of the repeaterless secret key capacity ($8.64\times10^{-7}$).
\end{abstract}

\maketitle

{\it Introduction.---}
Although quantum key distribution (QKD) can in principle offer secure private communication~\cite{BB84,Gisin2002,Dusek2006,Scarani2009,liao2017satellite,liao2018satellite,lydersen2010hacking}, there are still some technical limitations on practical long distance quantum communication. Perhaps the most severe of these is channel loss, given that quantum signals cannot be amplified. Much efforts have been made towards QKD over longer-distance. Theoretically, the decoy-state method~\cite{H03,wang05,LMC05} can improve the key rate of coherent-state based QKD from scaling quadratically to linearly with the channel transmittance, as what behaves of a perfect single-photon source. This method can defeat the photon-number-splitting attack to the imperfect source and the coherent state is used as if only the single-photon pulses were used for key distillation, and hence it can reach a key rate in the linear scale of channel loss, as the perfect single-photon source does.

Remarkable theoretical progress was made toward achieving practical, secure QKD over longer distance with the proposal of twin-field QKD~\cite{nature18}, which improves the key rate scaling to follow the square root of the channel transmittance. It shows that, the coherent-state source can actually be an advantage over the single-photon source because the post-selection of phase coherence of the twin fields from Alice and Bob can potentially lead to secure QKD with the encoding state of single-photon and vacuum, and their linear super-positions. This method has the potential to achieve a key rate that scales with the square root of channel transmittance, and can by far break the known distance limit of existing protocols in practical QKD~\cite{zhang2018large,MDI200km,braunstein2012side,lo2012measurement,WangPRA2016,MDI404km,boaron2018secure,takeoka2014fundamental,PLOB2017}. The theoretical secure key rate can be even higher than the repeaterless secret key capacities, known as the Takeoka-Guha-Wilde (TGW) bound~\cite{takeoka2014fundamental} and the
Pirandola-Laurenza-Ottaviani-Bianchi (PLOB) bound~\cite{PLOB2017}. However, considerable work still remains to make this a reality.

First, there is the theoretical challenge of combining the phase information post-selection with the traditional decoy-state method. Second, it is technically demanding of precise single photon interference over long distance. Towards this goal, the sending-or-not-sending (SNS) protocol~\cite{wang2018sns} was proposed. This involves small sending probabilities for both Alice and Bob, and then uses sending and not-sending decisions for the bit value encoding in $Z$-basis with the effective heralding events announced by Charlie. In this way, as was shown in~\cite{wang2018sns} one can continue to use the tagged model and the conventional decoy-state method in the protocol. In addition, since the protocol encodes the bit values using the almost error-free $Z$-basis, it can tolerate a high error rate in $X$-basis.

Here, we report an experimental demonstration of twin field QKD through SNS protocol (SNS-TF-QKD) over optical fiber spools.

\begin{figure}[tbh]
\centering
\resizebox{0.9\columnwidth}{!}{\includegraphics{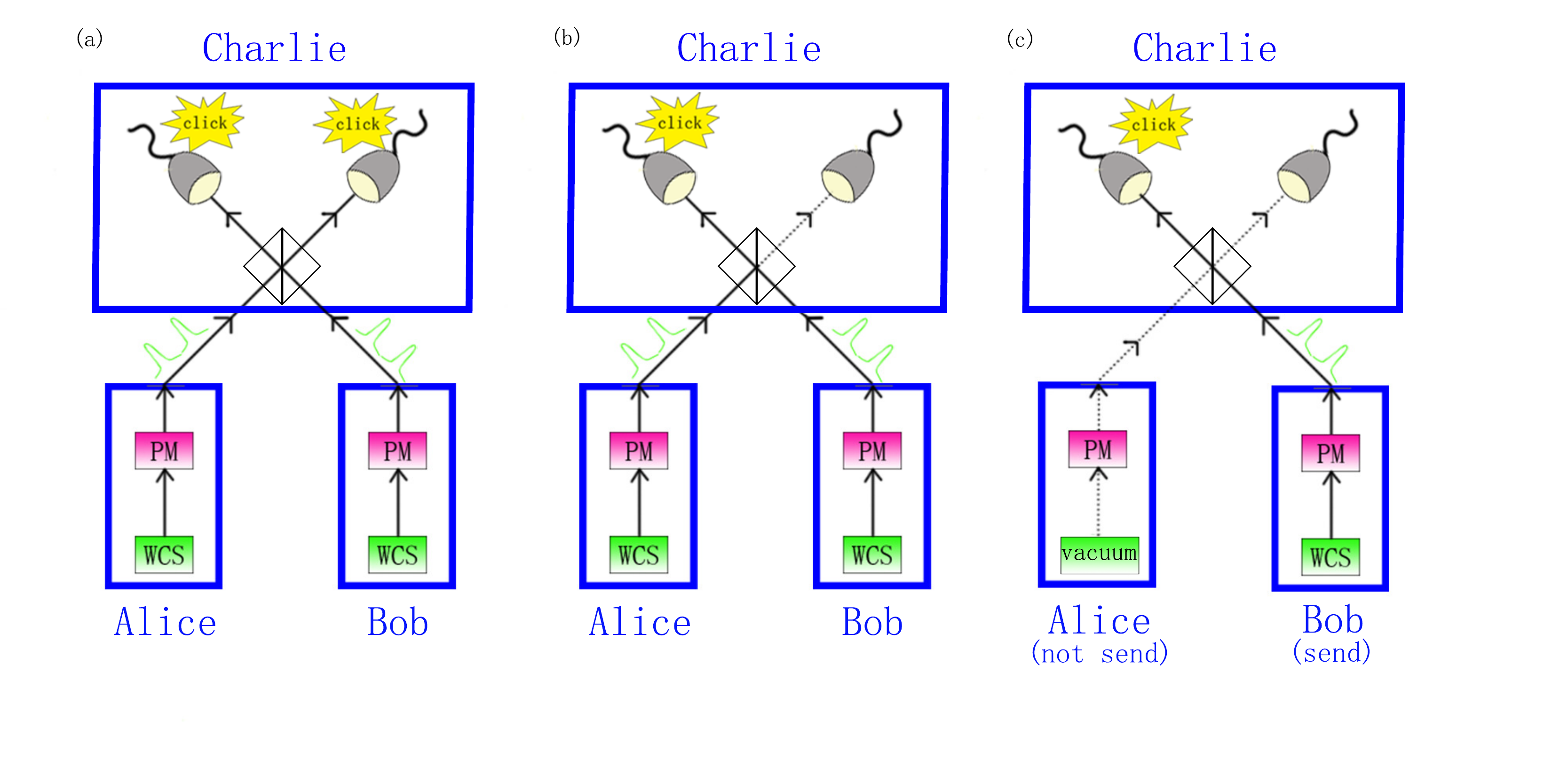}}
\caption{Schematics for three different protocols. 
(a) Decoy-state MDI-QKD, where coherent-state pulse pairs in BB84 encoding are sent out and effective events are heralded by two-fold clicking. The key rate scales linearly with the channel transmittance. 
(b) The original decoy-state TF-QKD~\cite{nature18}, where twin fields of coherent states with random phase shifts are sent out in the $X$ and $Y$-bases, and the effective events are heralded by single clicking. The key rate scales with the square-root of the channel transmittance. Single-photon interference from remote independent sources are needed for both bases. There can be misalignment errors in both bases and the post announced phase-shift information render the decoy-state method invalid.
(c) Decoy-state SNS-TF-QKD~\cite{wang2018sns}. In Z-basis, each side independently decide sending with a small probability. The events that one side decides sending, the other side decides not sending, and one and only one detector clicks (as shown in the figure) are the target events to generate secure keys. It is fault tolerant to large misalignment error in X-basis because there is no misalignment error in Z-basis. The traditional decoy-state method holds because the phase-shift information in Z-basis is never announced. The heralding of a single clicking makes the effective events in Z-basis and the key rate is in the scale of square root of channel transmittance. WCS: weak coherent source
}
\label{Fig:sketch}
\end{figure}

{\it Protocol.---}Consider the schematic of the SNS-TF-QKD protocol~\cite{wang2018sns} shown in Fig.~\ref{Fig:sketch}. Here, we implement the protocol using a practical four-intensity method~\cite{yu2018sending}, where each party exploits four different intensities, namely,  $0, \mu_1, \mu_2$ and $\mu_z$.

Alice and Bob randomly choose the $X$-or $Z$-basis with probabilities $p_X$ and $1-p_X$, respectively. In the $X$-basis, both Alice and Bob prepare and send decoy pulses. Phase shifts of $\theta_{A}$ and $\theta_B$ are privately applying to their pulses. A$Z$-basis event is regarded as being effective if Charlie announces that only one detector clicked. For an $X$-basis event to be effective, we need an extra phase-slice condition to reduce the observed error rate in the basis. Without a reasonable phase-slice condition, the observed $X$-basis error rate can be too much higher than the actual phase-flip error rate in the $Z$-basis. Note that Charlie does not have to be honest, and whatever he announces does not undermine the security. But if Charlie wants to make a good key rate, he will have to try to make a faithful announcement on everything.

An error in $X$-basis is defined as Charlie announcing a click of right (left) detector associated with an effective event in $X$-basis where the phase difference between the pulse pair from Alice and Bob would provably cause a left (right) clicking at Charlie's measurement set-up. A $Z$-basis effective event that Alice (Bob) has decided sending and Bob (Alice) has decided not-sending corresponds to a bit value 1 (0). The values of $e_1^{ph}$ and $s_1$, the yield of single-photon effective events in the $Z$-basis, can be calculated by the conventional decoy-state method~\cite{wang2018sns,yu2018sending}. See Supplemental Materials for detailed calculations.

\begin{figure*}[tbh]
\centering
\resizebox{0.95\columnwidth}{!}{\includegraphics{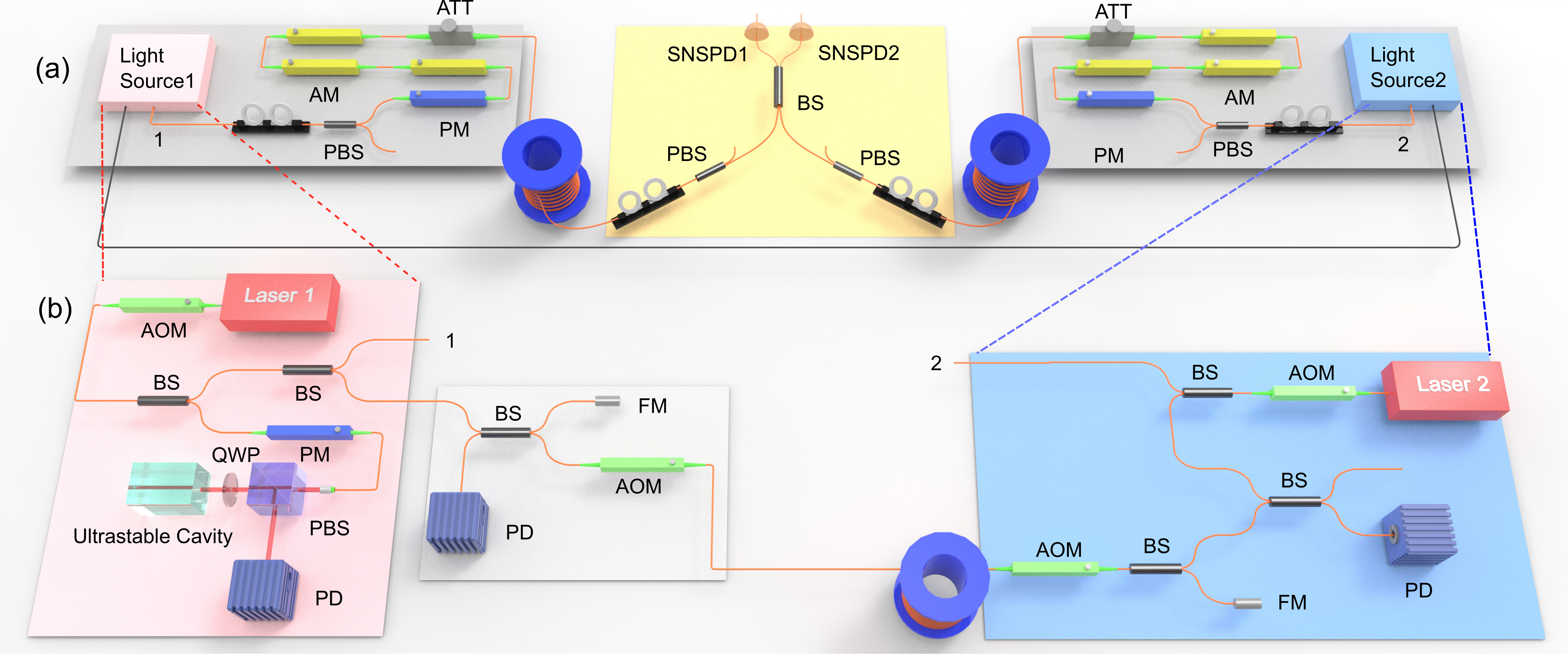}}
\caption{(a) Schematic of our experimental setup. Alice and Bob use frequency-locked continuous-wave (CW) lasers as sources. These lasers are then modulated by a phase modulator (PM) and three amplitude modulators (AMs) for phase randomization, encoding, and decoy intensity modulation. The pulses are then  attenuated by an attenuator (ATT) and sent out via fiber spools to Charlie. At Charlie's measurement station, the pulses from Alice and Bob pass through polarization controllers (PCs) and polarization beam splitters (PBSs), then interfere at a beam splitter (BS). Finally, the light is measured by superconducting nanowire single-photon detectors (SNSPDs). (b) Frequency-locking system for Alice's and Bob's lasers. The fiber length between Alice and Bob is set to be the same as the total length of the signal fiber. AOM: acousto-optic modulator, FM: Faraday mirror, PD: photodiode, QWP: Quarter-Wave Plate. }
\label{Fig:TFQKDSetup}
\end{figure*}

{\it Experiment.---}
The experimental layout is shown in Fig.~\ref{Fig:TFQKDSetup}(a). In Alice's and Bob's labs, independent continuous wave (CW) lasers are used as light sources. The light is modulated to 16 different phases with a phase modulator (PM) and encoded with three amplitude modulators (AMs). In the experiment, we set the basic period to 5 $\mu$s, during which, 100 signal pulses are sent in the first 3 $\mu$s with 2 ns pulse width and 30 ns interval, followed by 4 phase reference pulses in the next 1.2 $\mu$s, to estimate the relative phase between Alice's and Bob's channels, and with a final 0.8 $\mu$s vacuum state as the SNSPDs recovery time. The signal intensities are set to the optimized decoy states $\mu_z$, $\nu_1$, $\nu_2$ or $0$ (see Supplemental Materials for details about the encoding). 

Then, the signals are transmitted from Alice and Bob to Charlie, where they interfere. Since interference requires identical input signals, polarization controllers (PCs) and polarization beam splitters (PBSs)  before the polarization maintaining beam splitter (BS) are necessary to compensate the polarization drift of the channel. The interference results are then detected with SNSPDs and recorded with a high speed Time Tagger.

The main technical challenge in realizing SNS-TF-QKD is to control the twin fields' phase evolution. As was pointed out in~\cite{nature18}, the differential phase fluctuation between the two users can be written as:
\begin{equation}
	\delta_{ba}=\frac{2\pi}{s}(\Delta\nu L+\nu\Delta L)
	\label{eq:DPF}
\end{equation}
where $\nu$ is the optics frequency of the light, $L$ is the length of the fiber, $s$ is the speed of light in the fiber. Thus, two sources contribute to the phase difference should be compensated: the first term in the equation indicates the frequency difference between Alice and Bob; the second term stands for the phase drift in the fiber. As an example, the measured phase drift rate follows a Gaussian distribution with a standard deviation of 7.4 $rad\cdot ms^{-1}$ for 150 km total fiber distance. The drift a bit larger than reported in~\cite{nature18} mainly due to the relatively noisy environment in our lab (See Supplemental Materials for details).

To deal with phase difference caused by the wavelength difference, we adopt a frequency-locking method, as shown in Fig.~\ref{Fig:TFQKDSetup}(b). In Alice's lab, a continuous wave laser with a center wavelength of 1550.12 nm and a line width of several kilohertz is used as the seed laser. The seed laser is locked to a 10-cm-long ultra-stable cavity with a finesse around 250000, using the Pound-Drever-Hall (PDH) technique~\cite{drever1983laser,pound1946electronic}, to suppress its line-width from a few kilohertz to approximately ten hertz. The light is then split into two parts, one is used as Alice's source and another to lock Bob's optical frequency. This locking beam is further split into two parts, one is reflected by a Faraday mirror (FM) as the local reference, while the other is frequency-modulated by an acoustic-optic modulator (AOM) and sent to Bob. Here, the fiber length is set to be the same as the signal transmission distance to demonstrate the system's practicability.

To ensure Bob receives light with the same frequency as Alice's, we need to compensate for any phase noise in the transmission fiber. On Bob's side, a BS and an FM works as a partial reflector for the locking light. A second AOM with a fixed frequency shift is inserted before the BS, to distinguish the reflected laser light from backscattering and facet reflection noise. The whole setup, including Alice's and Bob's reflectors, is actually a large Michelson interferometer. Thus, the phase error signal can be generated by mixing the beat pattern at Alice's photodiode (PD) with a fixed-frequency reference. A PI servo then generates the feedback signal for Alice's AOM, compensating for the phase drift on Bob's side. Bob then locks his own laser to the transmitted one with fast AOM feedback and slow piezoelectric feedback with a large adjustment range. The whole locking system can work continuously for weeks.

To handle the phase difference caused by path fiber noise, we use phase reference pulses to estimate it. During the phase reference time, Alice adjust the phases to $\theta_A$=\{0, $\pi/2$, $\pi$ and 3$\pi$/2\} for the 4 type of pulses, while Bob maintains his phase at 0. So the total phase difference before the interference is $\phi=\theta_A-\theta_B+\Delta \varphi_T$, where $\Delta \varphi_T$ is the relative phase difference of the fiber channel. Thus, Charlie records the interference detections for four different phases $\phi$, and then uses these results to search for the most probable phase difference $\Delta \varphi_T$ using a least squares method (see Supplemental Materials for further details of the phase estimation process).

To accumulate enough detections to estimate the relative phase, we set the intensity of the phase reference pulses such that a total of approximately 20 detection events occurred in the 1.2 $\mu$s interval, which requires a peak counting rate of more than 40 MHz. This need for such a peak counting rate, together with low dark count noise of less than 1000 Hz, imposes stringent requirements on single photon detectors. In particular, the low dark count rate must be achieved within a recovery time of only a few hundred of nano-second, making the problem even harder.

We developed SNSPDs with two-parallel-nanowire serial-connected configuration~\cite{miki2017stable}, which cover an active area of 16 $\mu$m in diameter. The SNSPD's recovery time is mainly limited by its kinetic inductance~\cite{kerman2006kinetic}. This parallel configuration effectively reduces the kinetic inductance without scarifying detection efficiency. In addition, we inserted a 50 ohm shunt resistor between the DC arm of the bias tee and the ground at room temperature~\cite{liu2012nonlatching} to prevent the detector latching at high count rates.

\begin{figure}[tbh]
\centering
\resizebox{9cm}{!}
{\includegraphics{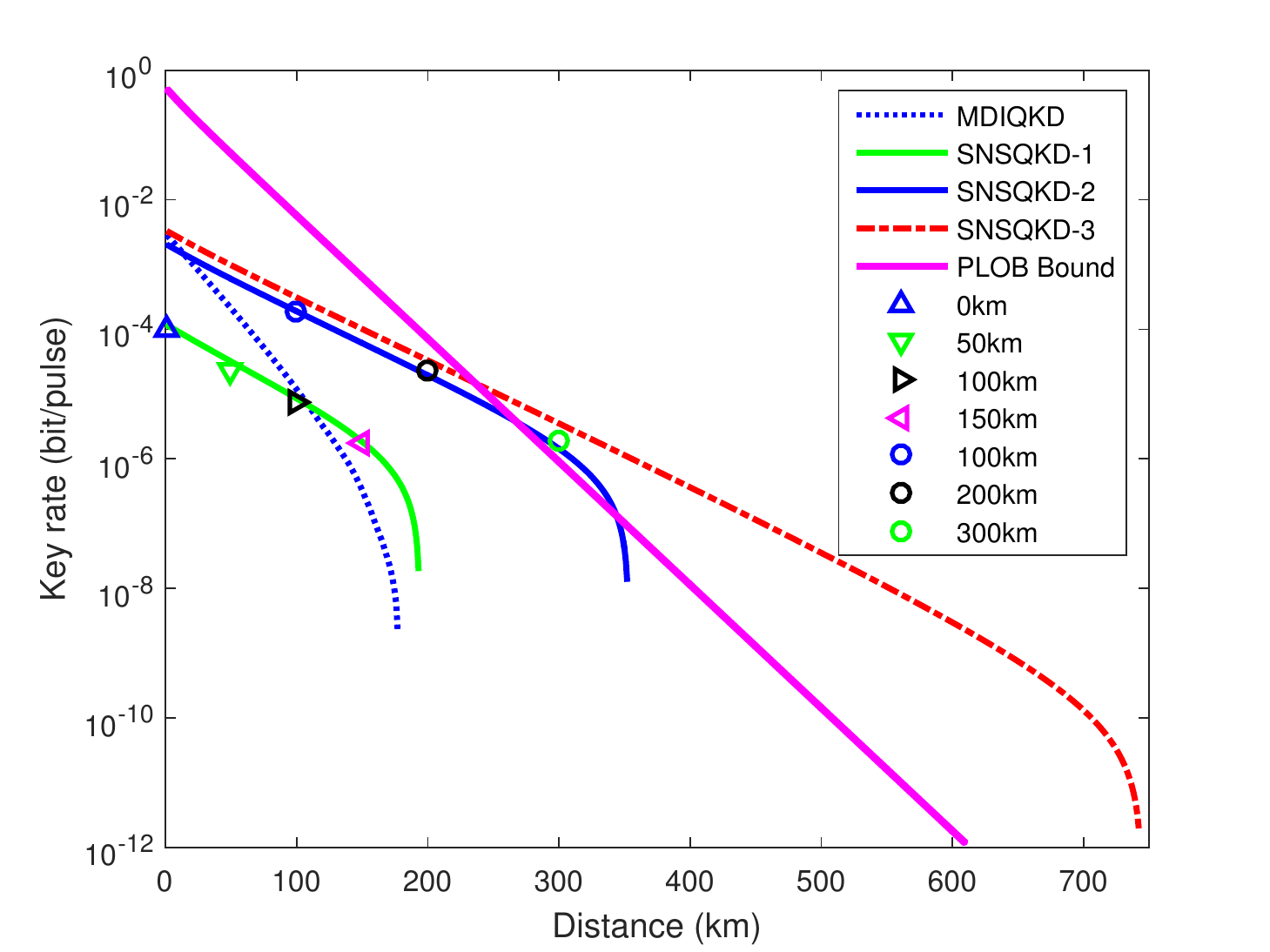}}
\caption{Secure key rates and SNS-TF-QKD simulation results. The triangles show the experimental results for the first experimental test while the solid green curve represents the simulation results with dark count probability of around $10^{-6}$ and X-basis baseline error of approximately 10\%; For comparison, the dotted blue curve gives the result of simulating four-intensity decoy-state MDI-QKD protocol~\cite{WangPRA2016} with the same parameters but with 2\% optical errors in the $X$-basis. The circles show the experimental results for the second test while the solid blue curve represents the simulation with dark count probability of around $10^{-7}$ and X-basis baseline error of approximately 2\%. The total pulses sent by Alice and Bob for all the experimental tests are $7.2\times10^{11}$. The dot-dashed red curve further assumes a total of $10^{14}$ pulses sent by Alice and Bob with 2\% X-basis baseline error. Finally, the solid magenta line illustrates the PLOB bound~\cite{PLOB2017}.}
\label{Fig:Result}
\end{figure}

Instead of actively stabilizing the relative phase between Alice and Bob, we compensate for the phase difference via post processing. Defining the estimated relative phase between Alice's and Bob's fibers as $\Delta \varphi_T$, we calculate the quantum bit error rate (QBER) in the $X$ basis for the detections that lie within the range
\begin{equation}
1-|\cos(\theta_A-\theta_B+\Delta\varphi_T)|<\Lambda
\end{equation}
where $\theta_A$ ($\theta_B$) is the random phase Alice (Bob) modulates on the signal and $\Lambda$ is a preset range. 

Then, we can calculate the secure key rate with finite data size effect by the following formula~\cite{jiang2019unconditional} (See Supplemental Materials for further details of the protocol):
\begin{equation}\label{eq:KeyRate}
\begin{split}
  R=(1-p_X)^2 \{ 2p_Z(1-p_Z) a_1 s_1 [1-H(e_1^{ph})]\\
  -f S_Z H(E_Z)\}-\frac{1}{N_{total}}\log_2\frac{1}{\epsilon^5},
\end{split}
\end{equation}
where $R$ is the final key rate, $a_1=\mu_z e^{-\mu_z}$, $s_1$ is the yield of effective single-photon events in the $Z$-basis, $e_1^{ph}$ is the phase-flip error rate for the events in $Z$-basis, $S_Z$ and $E_Z$ are the observed yield and bit-flip error rate for the $Z$-basis, $N_{total}$ is the total number of signal pulses sent, and $\epsilon=10^{-10}$, corresponding to a total failure probability of $2\times 10^{-9}$. The key rate would be even higher if we only considered the statistical fluctuation alone. Here, we assume an error correction efficiency of $f=1.1$. 

We tested SNS-TF-QKD with total fiber distances from 0 to 300 km between Alice and Bob. The detailed parameters are summarized in Supplemental Materials, including the optical efficiencies of the fibers and optical element, the intensities and proportions of the signal and decoy states for each fiber lengths. In all the experiments with different fiber lengths, the total number of pulses sent by Alice and Bob is set to $7.2\times 10^{11}$. The valid detections are $6.5\times10^9$, $2.3\times10^9$, $7.6\times10^8$, and $2.5\times10^8$ for 0 km, 50 km, 100 km and 150 km in the first experimental test; $1.7\times10^9$, $1.9\times10^8$ and $2.4\times10^7$ for 100 km, 200 km and 300 km in the second test. 

The experimental result is summarized in Fig.~\ref{Fig:Result}.
First, we experimentally tested SNS-TF-QKD at dark count probability of $10^{-6}$ (equivalent to 1000 Hz), X-basis error rate of 10\%. The secure key rate at 150 km is $1.72\times 10^{-6}$ per pulse, already higher than a simulated secure key rate of the MDI-QKD protocol~\cite{WangPRA2016} using the same parameter as in the experiment, but assuming lower (2\%) optical errors in the $X$-basis ~\cite{MDI404km}. In fact, the simulation shows that the secure key rate already exceeds that of MDI-QKD at 108 km. 

Next, we reduced the dark count probability to approximately $10^{-7}$ (equivalent to 100 Hz) by upgrading SNSPD to integrate an on-chip bandpass filter~\cite{yang2014superconducting} inside, and reduced the X-basis error rate to around 2\% by using linear amplifier to drive the modulators. The secure key rate at 300 km fiber distance is $1.96\times10^{-6}$, which is higher than the PLOB bound~\cite{PLOB2017} of $8.64\times10^{-7}$ per pulse. The simulation shows that SNS-TF-QKD breaks the bound at 267 km, and the transmission distance can be more than 350 km with the experimental parameters.

Finally, we simulate the secure key rate assuming a total of $10^{14}$ pulses are sent (with $2.6\times 10^5$ valid detections accumulated at 720 km), and the single photon detector dark count probability reduced to $10^{-11}$ (equivalent to 0.1 Hz with 100 ps pulse width)~\cite{yang2014superconducting}. All other parameters are set to that in the 300 km experiment. The simulation shows a maximum distribution distance of 742 km, and the SNS-TF-QKD protocol achieve a key rate above the PLOB bound~\cite{PLOB2017} when the fiber distance is greater than 236 km.

In conclusion, we have developed phase locking and phase compensation technologies, tested the SNS-TF-QKD protocol experimentally, and demonstrated the generation of secure keys at fiber distances of up to 300 km, yielding a higher key rate than the repeaterless secret key capacity. The key rate calculation has fully considered finite size effects, thus guaranteeing the security in practical situation. We note that both the distance and key rate can be significantly improved by using two-way classical communication \cite{xu2019general}. The experimental result also shows the SNS-TF-QKD protocol is robust against the phase mismatching, which is an important advantage in practice. The phase locking method used in the experiment has been found to be stable at a fiber distance of 1800 km~\cite{droste2013optical}, and the intensities of the phase reference pulses were within a few micro-watts even at 1000 km. With currently available technology and the results of theoretical simulations with practical parameters, we expect that distribution distances of more than 500 km will be achieved in the near future.

Note added.--Recently, we become aware of related works~\cite{minder2019experimental,wang2019beating,zhong2019proof}.

\emph{Acknowledgements.---}
The authors would like to thank L.-Y. Qu for assistance with frequency locking, C. Wu for help in preparing the figures, and C.-Z. Peng for the help in electronics.

\section*{Supplemental Material}

\section{Theory of SNS-TF-QKD protocol}
In this experiment, we implement the SNS-TF-QKD protocol~\cite{wang2018sns} using practical 4-intensity method~\cite{yu2018sending}.  In the 4-intensity decoy-state sending-or-not protocol, Alice and Bob randomly choose the $X$-basis (decoy pulses)  with probability $p_X$ and $Z$-basis (signal pulses) with probability $1-p_X$ to send or not to send a phase-randomized coherent pulse to an untrusted party, Charlie, who is expected to perform interference measurement. In $X$-basis, both Alice and Bob prepare and send the decoy pulses. Explicitly they randomly choose three sources $\rho_{\alpha_i}$ with probability $p_i$ for $i=0,1,2$, where $\rho_{\alpha_0}=\oprod{0}{0}$ is the vacuum source, $\rho_{\alpha_1}$ and $\rho_{\alpha_2}$ are two coherent sources with intensity $\mu_1$ and $\mu_2$ ($\mu_1<\mu_2$) respectively.  In $Z$-basis, Alice (Bob) randomly prepares and sends the coherent state $\rho_{\alpha_z}$ with probability $p_z$ and sends nothing else. The intensity of $\rho_{\alpha_z}$ is denoted by $\mu_z$. The coherent state whose phase is selected uniformly at random can be regard as a mixture of photon number states, i.e., $\rho_{\alpha_j}=\sum_k a_{k,j}\oprod{k}{k}$ with $a_{k,j}=e^{-\mu_j}\mu_j^{k}/k!$ for $j=0,1,2,z$.

Here we disregard those events with mismatched bases (Alice and Bob committed to different bases) or mismatched intensities (Alice and Bob committed to different intensities when both of them committed to X-basis). Charlie measures the incoming signals and records which detector clicks. When the quantum communication is complete, he publicly announces all the information about the detection event. Charlie's announcement of one and only one detector clicking makes an effective event if both of Alice and Bob committed to the $Z$-basis. If both Alice and Bob committed to the $X$-basis, in addition to Charlie's announcement of one and only one detector clicked, we also need the phase slice condition as discussed below.

Alice and Bob collect all the data with effective events and discard the rest. An encoding error in $Z$-basis is counted for an effective event corresponding to the situation that both of Alice and Bob had decided sending or both of them had decided not-sending. Since the sending or not-sending decisions correspond to the secret key, they cannot announce all of them for the $Z$-basis effective events. They can announce a random subset of them and deduce the error rate in Z-basis. The errors in $X$-basis are used to  estimate $e_1^{ph}$, the phase-flip error rate of those single-photon effective events in $Z$-basis.

Alice and Bob first announce the basis information ($X$-basis or $Z$-basis), then which source has been used and the phase information corresponding to the effective events where either of them choose the $X$-basis. With this information, Alice and Bob obtain the observable $N_{jk}(j,k=0,1,2,z)$, namely the number of instances when Alice and Bob have sent states $\rho_{\alpha_j}$ and $\rho_{\alpha_k}$, respectively. Correspondingly, the lowercases $n_{jk}$ are used to denote the number of effective events. The yields can be defined as $S_{jk}=n_{jk}/N_{jk}$.

As was noted in the theoretical paper~\cite{wang2018sns,yu2018sending}, in the case that the channel (Charlie) makes a perfect compensation, the global phase is removed and one can simply use $1-|\cos(\theta_A-\theta_B)|<\Lambda$ to expect a low observed error rate in $X$-basis.

Obviously, the security proof of the original SNS-TF-QKD protocol~\cite{wang2018sns,yu2018sending} actually allows a more general phase slice condition $1-|\cos(\theta_A-\theta_B+\Delta\varphi_T)|<\Lambda$, where $\Delta\varphi_T$ can be any value~\cite{jiang2019unconditional}. We define two sets $C_{\Delta^{+}}$ and $C_{\Delta^{-}}$ containing the instances where both Alice and Bob have sent $\rho_1$ in the $X$-basis with phase information $\theta_A$ and $\theta_B$ that fall within the slices $|\theta_A-\theta_B+\Delta\varphi_T|\leq D_s$ and $|\theta_A-\theta_B+\Delta\varphi_T-\pi|\leq D_s$ respectively, where $D_s$ is determined by $\Lambda$. Here $|x|$ means the degree of the minor angle enclosed by the two rays that enclose the rotational angle of degree $x$, e.g., $|-15\pi/8|=|15\pi/8| = \pi/8$, $|-\pi/10|= \pi/10$.

The number of instances in $C_{\Delta^{\pm}}$ are $N_{11}^{\Delta^{\pm}}=\frac{\Delta}{2\pi} N_{11}$. Here in this experiment, Charlie does not take active compensation and $\Delta\varphi_T$ is not zero. The number of effective events corresponding to $C_{\Delta^{\pm}}$ are denoted by $n_{11}^{\Delta_0^{\pm}}$ and $n_{11}^{\Delta_1^{\pm}}$ for detector 0 and detector 1 respectively.

In the real protocol, the number of total pulses send by Alice and Bob is finite. In order to extract a secure final key, we have to consider the effect of statistical fluctuations caused by the use of finite-sized key. Accordingly, with the observed values $S_{jk}(j,k=0,1,2,z)$ and the corresponding number of pulse pairs, one can lower bound the mean value $\mean{s_1^Z}$ by
\begin{equation}\label{eq:s1LM}
  \mean{s_1^Z}\geq \mean{\underline{s}_1^Z}=\frac{\mu_2^2 e^{\mu_1}\underline{S}_{1} -\mu_1^2 e^{\mu_2} \overline{S}_{2}-(\mu_2^2-\mu_1^2)\overline{S}_{00}}{\mu_1 \mu_2 (\mu_2-\mu_1)},
\end{equation}
and upper bound the mean value $\mean{e_1^{ph}}$ by
\begin{equation}\label{eq:e1phUM}
  \mean{e_1^{ph}}\leq \mean{\overline{e}_1^{ph}}=\frac{\overline{T}_{\Delta} -1/2e^{-2\mu_1} \underline{S}_{00}}{2\mu_1 e^{-2\mu_1} \mean{\underline{s}_1^Z}}.
\end{equation}
where $S_k=\frac{1}{2}(n_{0k}+n_{k0})/N_{0k}$ for $k=1,2$, $T_{\Delta}=\frac{1}{2}(n_{11}^{\Delta_1^{+}}/N_{11}^{\Delta^{+}} + n_{11}^{\Delta_0^{-}}/N_{11}^{\Delta^{-}})$, $N_{00}=p_0^2 N_X + 2 p_0(1-p_z) N_{XZ}$, $N_{01}=N_{10}=p_0 p_1 N_X + (1-p_z)p_1 N_{XZ}$, $N_{02}=N_{20}=p_0 p_2 N_X + (1-p_z)p_2 N_{XZ}$, and $\underline{\mathcal{U}}_{k}={\mathcal{U}}_{k}/(1+\delta_{k})$, $\overline{\mathcal{U}}_{k}={\mathcal{U}}_{k}/(1-\delta_{k}^{\prime})$, with $\mathcal{U}=S,T$ and $k=00,1,2,\Delta$.

By using the multiplicative form of the Chernoff bound, with a fixed failure probability $\epsilon$, we can give an interval of $\mean{\mathcal{U}_{k}}$ with the observable $\mathcal{U}_{k}$, $[\underline{\mathcal{U}}_{k}, \overline{\mathcal{U}}_{k}]$, which can bound the value of $\mean{\mathcal{U}_{k}}$ with a probability of at least $1-\epsilon$. Explicitly, with the function $\delta(x,y)=[-\ln (y/2)+ \sqrt{(\ln(y/2))^2-8 \ln (y/2)x}]/(2x)$, we have $\delta_{00}=\delta(N_{00}S_{00},\epsilon)$, $\delta_j=\delta((N_{0j}+N_{j0})S_j,\epsilon)$, and $\delta_{\Delta}=\delta((N_{11}^{\Delta^{+}}+N_{11}^{\Delta^{-}}) T_{\Delta},\epsilon)$. Then, we can use the mean values $\mean{\underline{s}_1^Z}$ and $\mean{\overline{e}_1^{ph}}$ defined in Eqs.(\ref{eq:s1LM}) and (\ref{eq:e1phUM}) to estimate the lower bound of the yield $\underline{s}_1$ and the upper bound of the phase-flip error rata $\overline{e}_1^{ph}$ as
\begin{equation}\label{eq:s1cL}
  \underline{s}_1=\mean{\underline{s}_1^Z}(1-\delta_1^c), \quad \overline{e}_1^{ph}=\mean{\overline{e}_1^{ph}}(1+\delta_1^{\prime c}),
\end{equation}
where $\delta_1^{c}=\delta(a_1 N_{zz}^{c} \mean{\underline{s}_1^Z}, \epsilon)$, $\delta_1^{\prime c}=\delta(a_1 N_{zz}^{c}\underline{s}_1 \mean{\overline{e}_1^{ph}}, \epsilon)$ with $N_{zz}^{c}=2p_z(1-p_z)N_{zz}$ and $a_1=\mu_z e^{-\mu_z}$ being the probability to emit a single-photon state from source $\rho_z$.

With the lower bound of $s_1$ and the upper bound of $e_1^{ph}$ in Eq.(\ref{eq:s1cL}), the final key rate can be calculated by~\cite{jiang2019unconditional}
\begin{equation}\label{eq:KeyRate}
\begin{split}
  R&=(1-p_X)^2 \{ 2p_z(1-p_z) a_1 s_1 [1-H(e_1^{ph})]\\
  &-f S_Z H(E_Z)\}-\frac{1}{N_{total}}(\log_2\frac{2}{\varepsilon_{cor}} +2\log_2\frac{1}{\sqrt2\varepsilon_{PA}\hat{\varepsilon}})
\end{split}
\end{equation}
where $R$ is the final key rate, $a_1=\mu_z e^{-\mu_z}$, $s_1$ is the yield of effective single-photon events in the $Z$-basis, $e_1^{ph}$ is  the phase-flip error rate of those single-photon effective events in $Z$-basis, $S_Z$ and $E_Z$ are the observed yield and bit-flip error rate for the $Z$-basis, $N_{total}$ is the total number of signal pulses sent, and $f$ is the error correction efficiency factor.

With this key rate, the protocol is $\varepsilon_{sec}$-secret and $\varepsilon_{cor}$-correct. The security coefficient of the whole protocol is $\varepsilon_{tol}=\varepsilon_{sec}+\varepsilon_{cor}$, where $\varepsilon_{sec}=2\hat\varepsilon+4\bar\varepsilon+\varepsilon_{PA}+\varepsilon_{s_1}$. Here, $\varepsilon_{cor}$ is the failure probability of error correction; $\varepsilon_{sec}$ is the probability that the final key isn't secure;  $\bar{\varepsilon}$ is the accuracy of estimating the smooth min-entropy, which is also the failure probability that the real value of $e_1^{ph}$ isn't in the bound that we estimate; $\varepsilon_{PA}$ is the failure probability of privacy amplification; $\varepsilon_{s_1}$ is the failure probability that the real value of $s_1$ isn't in the bound that we estimate.
Here we set $\varepsilon_{cor}=\hat{\varepsilon}=\varepsilon_{PA}=\epsilon= 10^{-10},\bar{\varepsilon}=3\epsilon$ and $\varepsilon_{s_1}=4\epsilon$, and thus security coefficient of the whole protocol is $\varepsilon_{tol}=20\epsilon=2\times 10^{-9}$. The reason we set $\bar{\varepsilon}=3\epsilon$ and $\varepsilon_{s_1}=4\epsilon$ is that we use the Chernoff bound for three times to estimate $e_1^{ph}$ and four times to estimate $s_1$, respectively.

\section{Controlling the Relative Phase between Alice and Bob}
\subsection{Relative Phase Drift between Alice and Bob}
The biggest challenge in SNS-TF-QKD experiment is to control the relative phase between Alice and Bob. As was pointed out in~\cite{nature18}, the differential phase fluctuation between the two users can be written as

\begin{equation}
	\delta_{ba}=\frac{2\pi}{s}(\Delta\nu L+\nu\Delta L)
	\label{eq:DPF}
\end{equation}
where $\nu$ is the optical frequency of the light, $L$ is the fiber length, and $s$ is the speed of light in the fibre. Here, the first term represents the phase fluctuations caused by the phase difference between Alice and Bob, while the second represents the phase fluctuations caused by the changes in fiber path between Alice and Bob.

Previous reports~\cite{nature18} found that the phase drifts between fiber spools were determined to 2.4 $rad\cdot ms^{-1}$ and 6.0 $rad\cdot ms^{-1}$ at total distances of 100 and 550 km, respectively. These results indicate that it is possible to compensate for the phase fluctuation due to the fiber path. In this section, we focus on the phase fluctuation due to the first term in Eq.~\ref{eq:DPF}, namely, the light source.

The light source may affect the phase difference for two reasons: The wavelengths and thus the relative phase of Alice's and Bob's laser sources may be different and varying. In words, these frequency difference may cause beats. To provide feedback, we need to rapidly measure the relative frequency before the phase drifts too much due to this frequency beating effect. From Eq.~\ref{eq:DPF}, if we consider the phase fluctuations of $\delta_{ba}=0.01$ for $L=$2 km, which is equivalent to 10 $\mu$s, we can easily calculate that the optical frequency difference should be $\Delta\nu<$159 Hz. Thus we require that the frequency differences between the two laser sources, and for the same source at different time, are less than $\Delta\nu<$159 Hz.

\subsection{Measuring the Phase Drift due to  Frequency Differences in the Sources}
Here, we estimate the effect of the phase drift by measuring it using independent laser sources, with fiber spools in between the sources and the measurement station.

\begin{figure}[htb]
  \centering
  \resizebox{8cm}{!}{\includegraphics[scale=1]{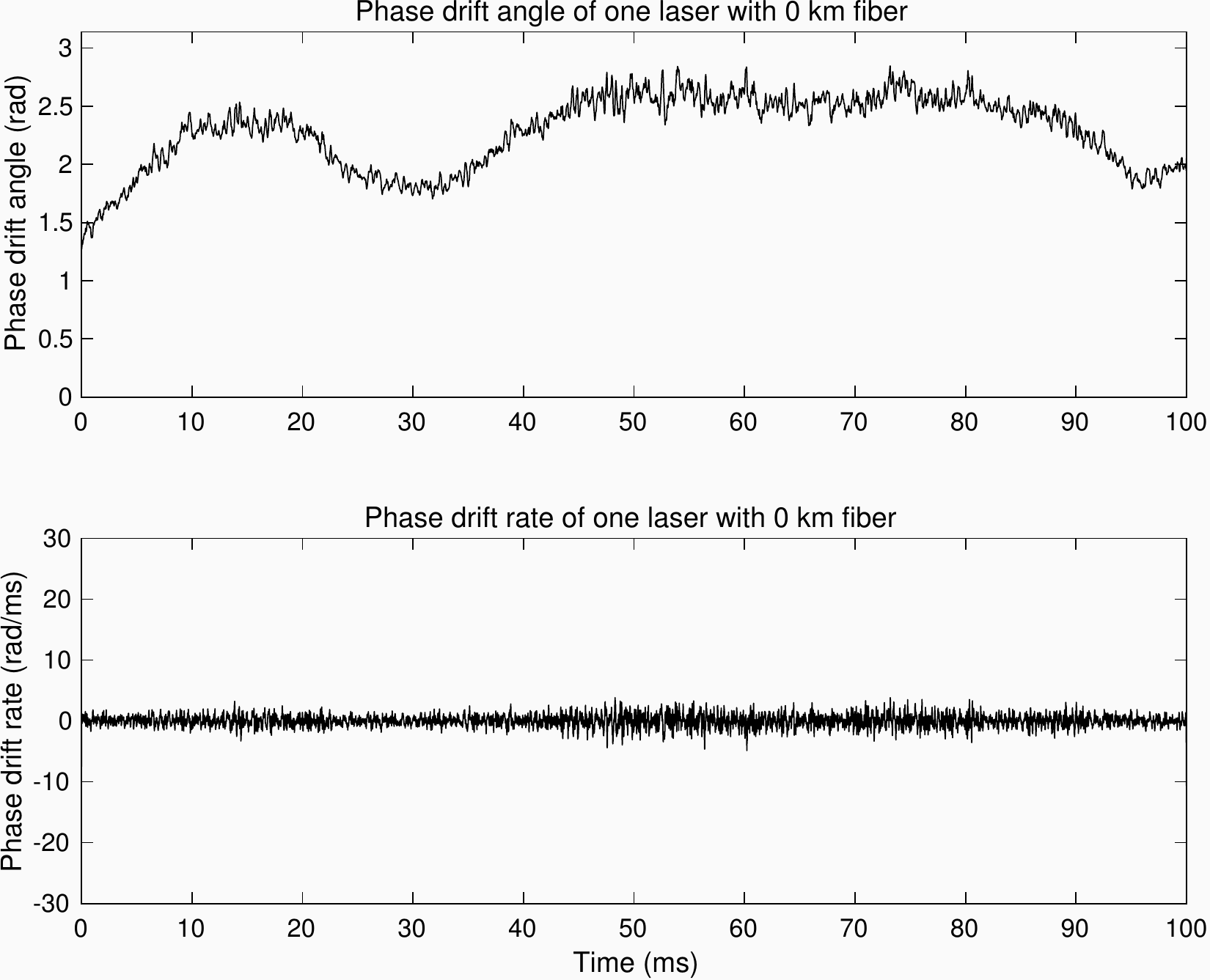}}\\
  \caption{Measurement of the fiber drift of one laser source with 0 km fiber in between.}\label{Fig:PhaseDrift0km1source}
\end{figure}

First, we use a single laser as the source, splitting its beam into two paths, and then combining them with a beam splitter (BS) for interference. For comparison with later experiments, this scenario did not use fiber spools. The setup is actually just a simple balanced Mach-Zehnder (MZ) interferometer. The  phase drift angle and the phase drift rate results are shown in Fig.~\ref{Fig:PhaseDrift0km1source}. Here, the measured phase drift rate follows a Gaussian distribution with a standard deviation of 1.0 $rad\cdot ms^{-1}$.

\begin{figure}[htb]
  \centering
  \resizebox{8cm}{!}{\includegraphics[scale=1]{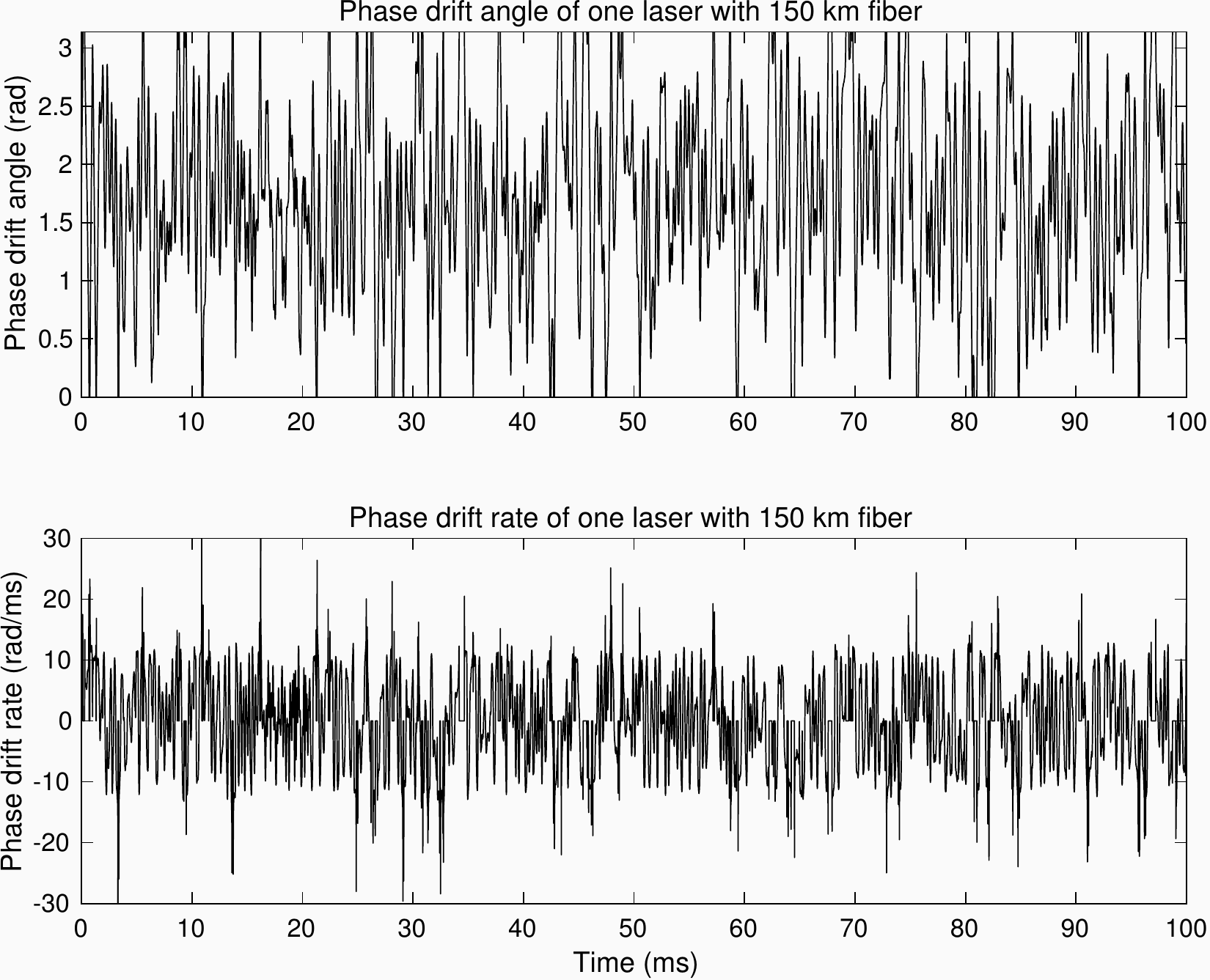}}\\
  \caption{Measurement of the fiber drift of one laser source with 150 km fiber in between.}\label{Fig:PhaseDrift150km1source}
\end{figure}

Second, we inserted 75 km fiber spools into each arm of the MZ interferometer. In other words, the longest fiber distance of 150 km between Alice and Bob is tested in this scenario. The phase drift angle and the phase drift rate results are shown in Fig.~\ref{Fig:PhaseDrift150km1source}. The measured phase drift rate follows a Gaussian distribution with a standard deviation of 7.1 $rad\cdot ms^{-1}$.

\begin{figure}[htb]
  \centering
  \resizebox{8cm}{!}{\includegraphics[scale=1]{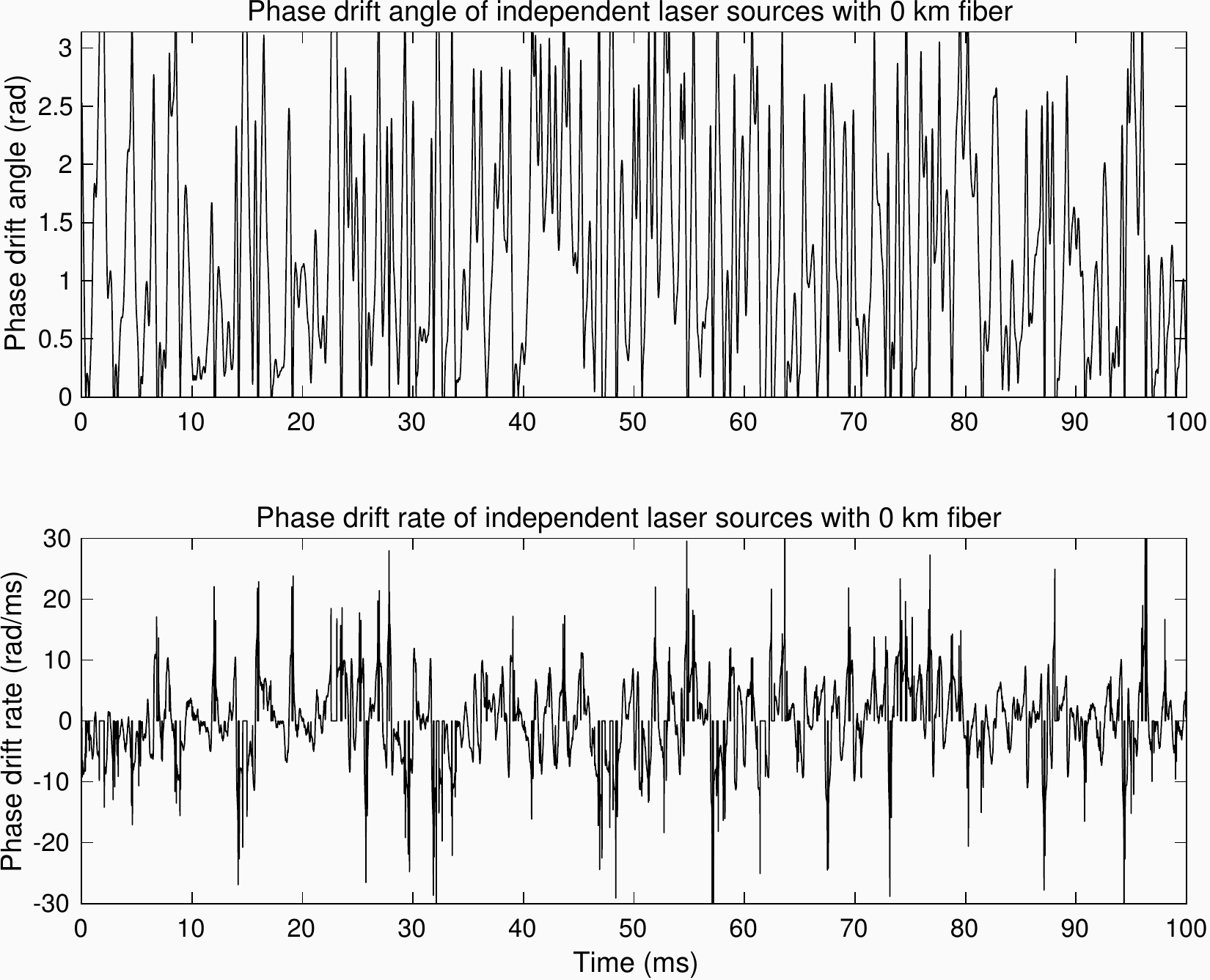}}\\
  \caption{Measurement of the fiber drift of two laser sources locking with each other with 0 km fiber in between.}\label{Fig:PhaseDrift0km2source}
\end{figure}

Third, we used two phase-locked independent lasers as the source. The light is directly interfered at a BS, as in scenario I. This scenario tested the performance of the phase-locked lasers. The phase drift angle and phase drift rate results are shown in Fig.~\ref{Fig:PhaseDrift0km2source}. The measured phase drift rate follows a Gaussian distribution with a standard deviation of 5.8 $rad\cdot ms^{-1}$.

\begin{figure}[htb]
  \centering
  \resizebox{8cm}{!}{\includegraphics[scale=1]{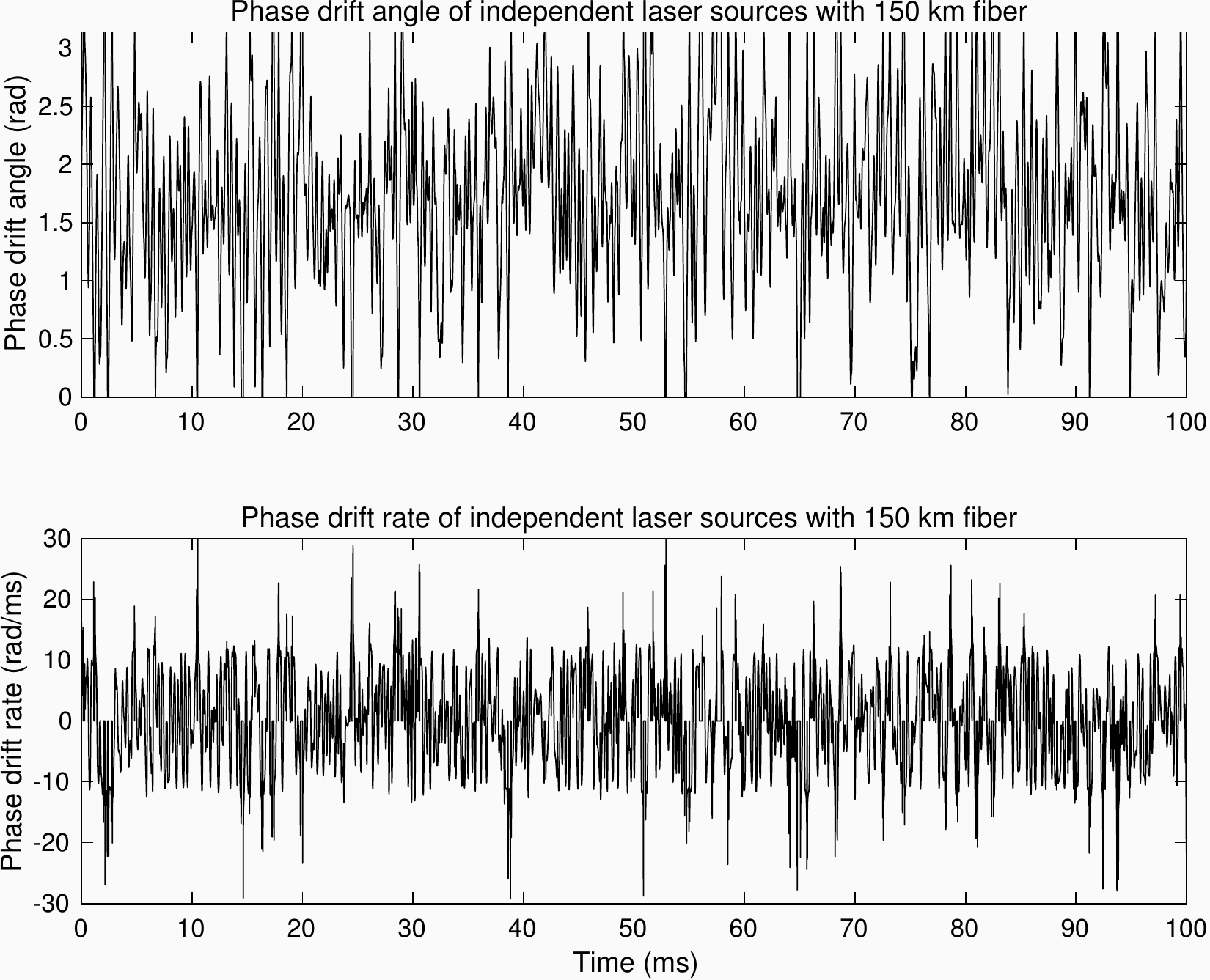}}\\
  \caption{Measurement of the fiber drift of two laser sources locking with each other with 150 km fiber in between.}\label{Fig:PhaseDrift150km2source}
\end{figure}

Finally, we inserted 75 km fiber spools in between the sources and the measurement. This tested independent phase-locked lasers and the long fiber distance (150 km) between Alice and Bob. The phase drift angle and phase drift rate results are shown in Fig.~\ref{Fig:PhaseDrift150km2source}. The measured phase drift rate follows a Gaussian distribution with a standard deviation of 7.4 $rad\cdot ms^{-1}$.

During the test, the standard deviation of the drift rate is less than 7.4 $rad\cdot ms^{-1}$ for all scenarios, including with independent sources, and the phase drift induced by 150 km fiber spools. The maximum drift rate is 31 $rad\cdot ms^{-1}$, which is around 0.31 rad (or 17 degree) in 10 $\mu$s. The phase drift related error is then less than 3\% when Alice and Bob sent the same phase. We set this 10 $\mu$s value as the phase reference read out period, with acceptable errors given by the phase drift during this period.

\subsection{Detailed Encoding Method in Experiment}
In the experimental setup as discussed in the main text, independent continuous wave (CW) lasers are used as light sources in Alice's and Bob's labs. The light is first modulated to 16 different phases with a phase modulator (PM) and then encoded with three intensity modulators (IMs). IM1 modulates the pulse intensity to either the signal state $\mu_z$, one of the decoy states $\nu_1$, $\nu_2$ or the vacuum state $0$; IM2 modulates the intensities of the phase reference pulses and signal state to the designed ratio so the output signal intensity is set to appropriate level and the reference detections are high enough for phase compensation; IM3 further modulates the signal pulse width to 2 ns width. All the modulators block the light when the pulse is a ``vacuum'' state, or is a ``not-sending'' state in the $Z$-basis. The signal intensities are further attenuated to the single-photon level by an attenuator before they leave Alice's or Bob's secure zone. 

In Alice's and Bob's labs, the pulse modulation is controlled by an arbitrary-wave generator (AWG) using pre-generated quantum random numbers. In the experiment, we used 5 $\mu$s as a basic period, during which, 100 signal pulses are sent in the first 3 $\mu$s with 30 ns intervals, followed by 4 type of phase reference pulses in the next 1.2 $\mu$s, to estimate the relative phase between Alice's and Bob's channels, with a final 0.8 $\mu$s vacuum state as the SNSPDs recovery time. The sampling rate of the AWG is 2 GHz with 14-bit depth, the waveform is then constructed based on the encoding information for each signal and reference pulses.

\subsection{Estimating the Relative Phase Drifts in Alice's and Bob's Fibers}
In this experiment, instead of compensating for the phase drift in fiber using a strong reference in real-time, we estimated the relative phase between Alice and Bob with the reference light, then corrected the relative phase in post-processing. With this post-processing method, we do not need fast electronics for real-time phase correction. We will discuss the details in the follow two subsections.

\begin{figure}[htb]
  \centering
  \resizebox{9cm}{!}{\includegraphics[scale=1]{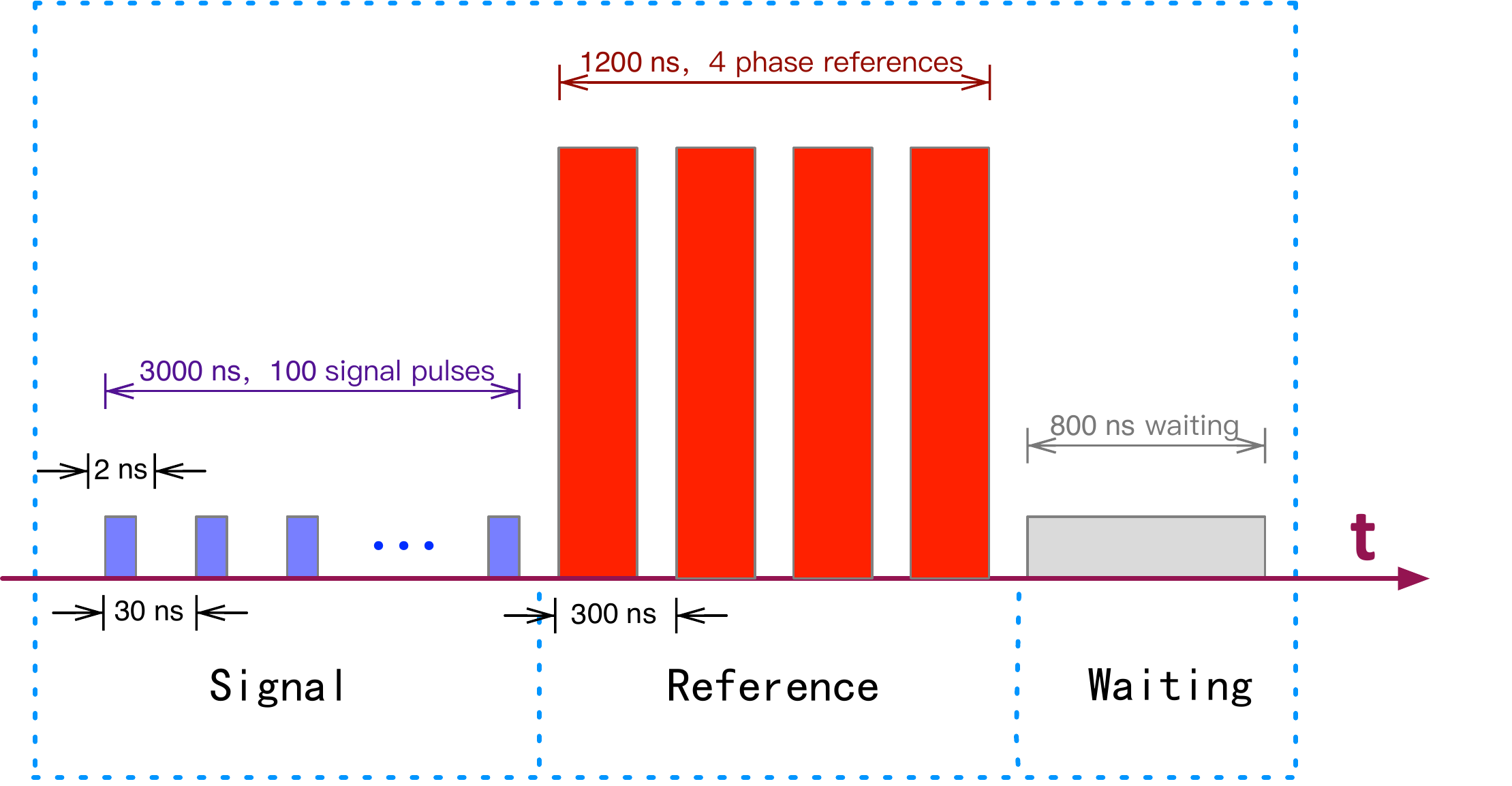}}\\
  \caption{Signal and reference pulses sent during one 5 $\mu$s period.}\label{Fig:SystemPeriod}
\end{figure}

The first step in controlling the relative phase between Alice's and Bob's fiber spools is to monitor and estimate it. For this we designed the signal and reference pulse pattens shown in Fig.~\ref{Fig:SystemPeriod}. During each 5 $\mu$s time period, we sent 100 signal pulses at a system frequency of 33.3 MHz, i.e., with a period 30 ns. Alice and Bob then modulated these pulses' phases and the intensities with phase and intensity modulators. Following each 3 $\mu$s signal pulses, we sent 4 reference pulses, representing different phases over the next 1.2 $\mu$s, allowing 300 ns for each phase. Alice modulated these phase to 0, $\pi/2$, $\pi$ and $3\pi/2$ in turn, while Bob maintained the phase at $0$. During each 300 ns, the specific pulse shape may be modulated for both Alice and Bob, to adjust the peak intensity and to adapt to the frequency response of the amplifiers and modulators. We used these phase reference pulses to monitor the relative phase drift between Alice's and Bob's fibers, with the detection result at Charlie. Finally, after the strong reference pulses, 800 ns is used as the recovery time of the superconducting nanowire single photon detectors (SNSPDs), when both Alice and Bob modulate the pulses to vacuum. We estimated our detectors would recover to the low dark count regime after this time.

Next, the light from Alice and Bob were transmitted through the fiber, interfering when they reach Charlie. Here, the interference at the BS's first output port is:
\begin{equation}
\begin{aligned}
I(\phi)&=\vert1+\mathrm{e}^{\mathrm{i}\phi}\vert^2\\
&=2 [1+\cos(\phi)]\\
&=4 \cos^2(\phi/2)
\end{aligned}
\end{equation}
where the phase $\phi$ represents the total phase difference between Alice and Bob. If the phase modulated by Alice (Bob) is $\theta_A$ ($\theta_B$), then the relative phase difference between them is $\Delta\theta=\theta_A-\theta_B$. If we also assume that the phase change in the fiber between Alice (Bob) and Charlie is $\varphi_A$ ($\varphi_B$), then the relative phase difference due to the fiber links is $\Delta\varphi_T=\varphi_A-\varphi_B$. This gives us total relative phase of:
\begin{equation}
\phi=\theta_A-\theta_B+\Delta\varphi_T
\end{equation}
and a normalized intensity from Charlie's interference measurement, given this total relative phase, of:
\begin{equation}
\begin{aligned}
I(\phi)&=[1+\cos(\phi)]/2\\
&=\cos^2(\phi/2)
\end{aligned}
\end{equation}
For single photon detection, the normalized intensity at Charlie represents the probability of him detecting a signal at his BS's first output port.

Based on these result, our design aims to estimate the phase difference between Alice's and Bob's fibers $\Delta\varphi_T=\varphi_A-\varphi_B$, given the phase reference pulses as input. As descried above, Alice adjust the reference pulses to four different phases, while Bob does nothing. Thus, there are four type of phase reference pulse with phase differences of: $0$, $\pi/2$, $\pi$, and $3\pi/2$.

During each 5 $\mu$s period, we counted the number of times the phase difference was detected as $N_0$, $N_{\pi/2}$, $N_{\pi}$ and $N_{3\pi/2}$. For accuracy, these were summed over more than one period.The phase reference pulses'  intensities were adjusted to around 2 MHz for each detector, and the detections were counted over 10 $\mu$s (2 periods). Then, we calculated the normalized counts, or probabilities, as
\begin{equation}
p_{i}=2 N_{i}/\Sigma N_{i}
\label{Eq:Prop}
\end{equation}
where $i=1...4$ indicates the different phase difference between Alice and Bob, namely
\{$0$, $\pi/2$, $\pi$, $3\pi/2$\}.

Next, we built the following error model:
\begin{equation}
Err(\Delta\varphi_T) = \sum_i{[p_i-p_{Ti}(\Delta\varphi_T)]^2}
\label{Eq:Err}
\end{equation}
where again $i=1...4$ stands for the four phases differences, and $p_{Ti}(\Delta\varphi_T)$ is the theoretical detection probability when the relative phase between the two fibers is $\Delta\varphi_T$. These theoretical probabilities are calculated as:
\begin{equation}
p_{Ti}(\Delta\varphi_T)=\cos^2(\frac{\Delta\theta_i+\Delta\varphi_T}{2})
\end{equation}
where the phase differences are $\Delta\theta_i=$\{$0$, $\pi/2$, $\pi$, $3\pi/2$\}, for $i=1...4$.

The final step is to minimize $Err(\Delta\varphi_T)$, as this is minimized when $\Delta\varphi_T$ matches the practical value. We adjusted $\Delta\varphi_T$ from 0$^\circ$ to 360$^\circ$, in steps of 1$^\circ$. Recording the phase $\Delta\varphi_{Tmin}$ that minimized $Err(\Delta\varphi_T)$ as the estimated phase difference. We also recorded the maximum value of $Err(\Delta\varphi_T)$, and defined the a parameter $rc$ as
\begin{equation}
rc = \min[Err(\Delta\varphi_T)]/\max[Err(\Delta\varphi_T)]
\end{equation}

This parameter indicates how close the $p_i$ distribution is to the theoretical one. Since the value $rc$ measures the ratio between the minimum and maximum errors, it approaches zero when the measured distribution is perfect, and one when the counts for all four phase reference pulses are the same. We used this parameter as a measure of the correctness of the estimate. 

In the experiment, Charlie is allowed to filter the detections based on his knowledge. The security holds as this operation is applied uniformly to all detections, regardless of whether Alice (Bob) sends in the $Z$- or $X$-basis. We set two criterion in our tests. The first is to apply a digital window to select the signal in the middle of each pulse, where the interference is expected to be better. The fraction of the data falling within this window is represented by the parameter $r_{gate}$. The second is to set a bound only to keep the detections where the reference pulses' $rc$ values were smaller than the pre-set bound. We note that the inaccurate phase estimation can be caused by high noise levels, or due to insufficient number of phase reference detections. The bounds can be adjusted to optimize the secure key rate.



\begin{table}[htb]
\centering
  \caption{Measured reference counts and probabilities, recorded over 10 $\mu$s. The estimated fibre phase difference is $\Delta\varphi_T=209^\circ$, with success probability of $rc=0.01$. This table also shows the theoretical probabilities calculated for $\Delta\varphi_T=209^\circ$. }
\begin{tabular}{ccccc}
\hline
Relative Phase $\Delta\theta$ & $0$ & $\pi/2$ & $\pi$ & $3\pi/2$\\
\hline
Measured Count & 4.15 & 16.60  & 22.55  & 6.25 \\
Measured Probabilities & 0.168 & 0.670  & 0.910  & 0.252 \\
Theoretical Probabilities & 0.063 & 0.742  & 0.937  & 0.258 \\
\hline
\end{tabular}
\label{Tab:PhaseEstimation}
\end{table}

\begin{figure}[htb]
  \centering
  \resizebox{6cm}{!}{\includegraphics[scale=1]{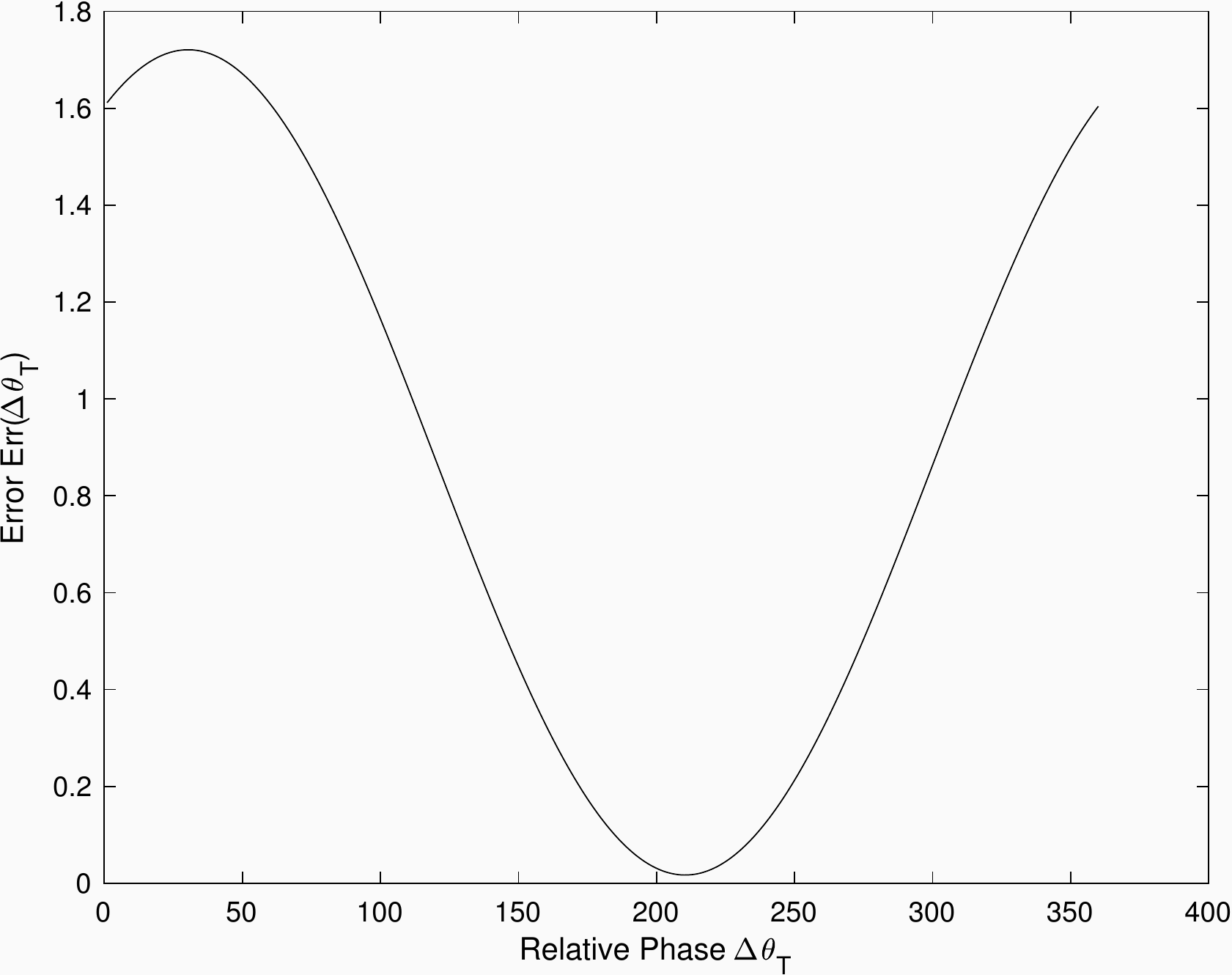}}\\
  \caption{Calculated error $Err(\Delta\varphi_T)$ verses the phase difference $\Delta\varphi_T$.}
  \label{Fig:ErrCurve}
\end{figure}

Tab.~\ref{Tab:PhaseEstimation} shows some example results. ``Measured Counts'' stands for the detections in 10 $\mu s$ for the four different phases, ``Measured Probabilities'' are the normalized probabilities calculated with Eq.~\ref{Eq:Prop}. Note we assumed a 40 ns dead time and the ``Measured Counts'' were recovered by accounting for the detection efficiencies and dead time. Next, to obtain the estimated phase difference, we optimize the error using Eq.~\ref{Eq:Err} varying $\Delta\varphi_T$ from 0 to 360 to obtain the result is shown in Fig.~\ref{Fig:ErrCurve}. From this result, we can read the most possible phase difference is $\Delta\varphi_T=209^\circ$, the phase estimation success probability $rc=0.01$.

\begin{figure}[htb]
  \centering
  \resizebox{7cm}{!}{\includegraphics[scale=1]{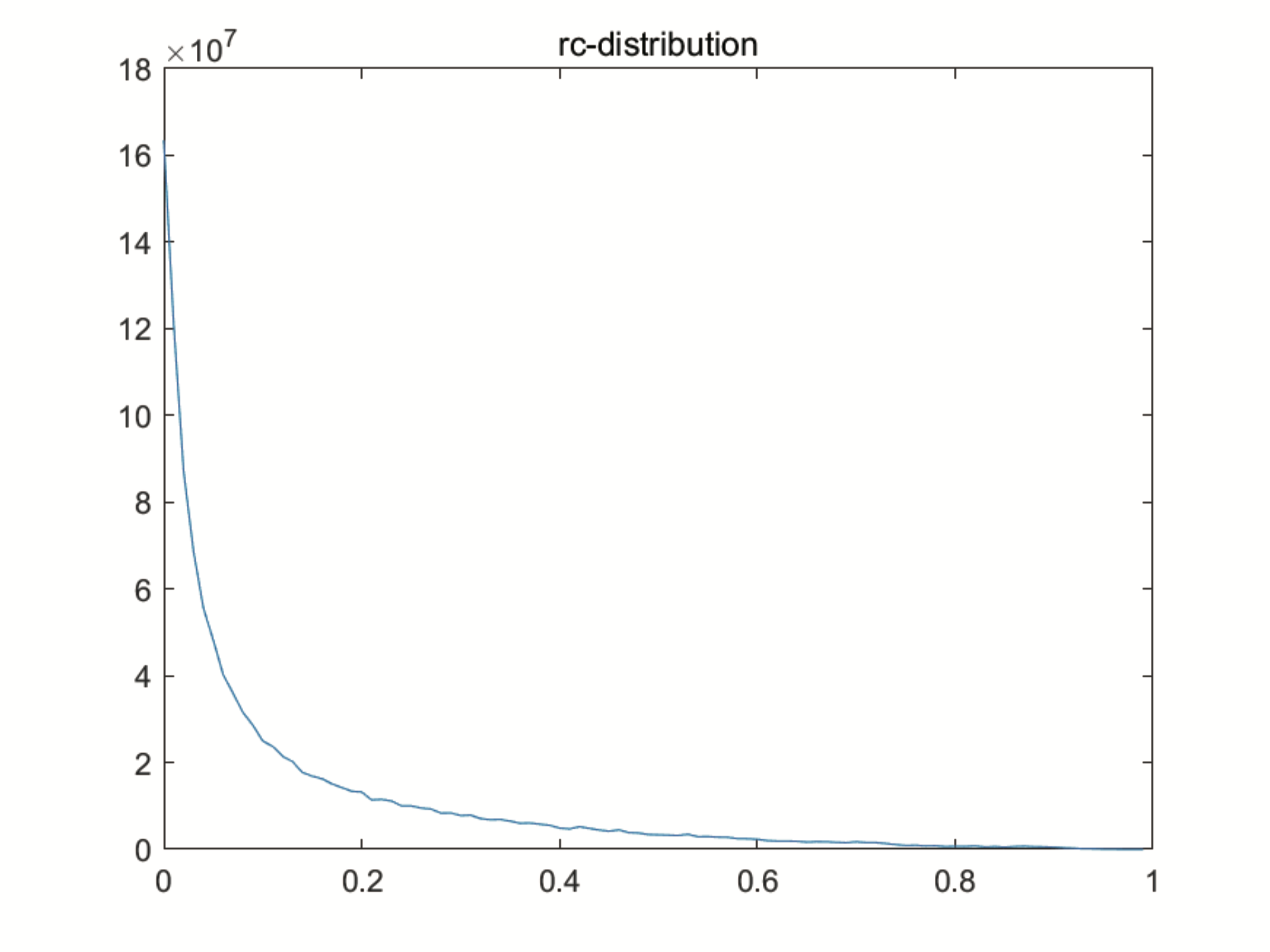}}\\
  \caption{Distribution of $rc$ based on the statistics for 100 km of fiber.}
  \label{Fig:RCDist}
\end{figure}

The phase estimation success probability $rc$ is different for different time periods, with different environments and noise levels. Take the data measured in 100 km as an example, we calculate the $rc$ distribution for the time period where there are detections, as shown in Fig.~\ref{Fig:RCDist}, optimizing the $rc$ values to maximize the final key rate for each fiber distance. In our experiments, a value of $rc=0.01$, with enabled around $14\%$ of the data to be kept, was the best choice. But this value will vary depending on the channel conditions.

\subsection{Compensating for the fiber-induced relative phase via post-processing}
The theoretical analysis assumes that the fiber-induced relative phase ($\Delta\varphi_T$) can always be stabilized to 0, a scenario that can indeed be realized with a strong reference light and a fast feedback circuit. In our experiment, however, we instead estimated the relative phase, as described above. We now demonstrate how we compensated for this relative phase,

In SNS-TF-QKD, we need to set a reasonable phase slice criterion for post-selecting $X$-basis events in order to derive a reasonable estimate of the $X$-basis  error rate. In our experiment, the channel (Charlie) does not apply the active compensation, and we instead use the following post-selection criterion:

\begin{equation}
1-|\cos(\theta_A-\theta_B+\Delta\varphi_T)|<\Lambda
\end{equation}
based on the estimated relative phase $\Delta\varphi_T$.
The original equation requires the phase difference between Alice and Bob to be less than $D_s$ or $D_s+\pi$. Using this revised equation, the phase difference after the fiber $\vert\theta_A-\theta_B + \Delta\varphi_T\vert$, falls into this range, i.e.,  $\vert\theta_A-\theta_B + \Delta\varphi_T\vert<D_s$ or $\vert\theta_A-\theta_B + \Delta\varphi_T\vert<D_s+\pi$, with high probability. Thus, the error due to the interference can be small.

In summary, we use phase reference pulse to estimate the phase difference between the two fibers. This method only requires an intensity of less than 50 total detections over 10 $\mu$s, meaning that the required laser source intensity within a reasonable range even when the communicating over several hundred kilometers. Using relatively weak reference pulses also reduces noises, which is a severe problem for long-distance communication. In addition, this approach does not require a fast feedback circuit, simplifying the system.

\section{Detailed Experimental Parameters}
Tab.~\ref{Tab:Parameters} summarizes the parameters used in our experiment. The lengths of the fibers between Alice (Bob) and Charlie are the same, and the ``fiber length'' in the table is the total fiber length, $\mu_1$, $\mu_2$ and $\mu_z$ are the intensities for the decoy states and the signal state. Here, the intensity of the weak decoy state $\mu_1$ was fixed at a mean photon number of 0.05. We did not use the (smaller) optimized value, mainly because of the intensity modulator performance limitations: stable intensity modulation is required for the decoy state, but the modulator is very sensitive and varied over time if the extinction ratio is too high. Thus, we used the lowest intensity that still ensured the decoy intensities are stable.

The phase reference pulses had the same width as the signal pulses (2 ns), and their intensity, denoted by $\mu_{ref}$, is given as the intensity during the time the reference pulses were not modulated to vacuum. The ``reference width'' indicate the interval during which Alice (Bob) sent reference pulses in each 300 ns. We did not send pulses for the full 300 ns for two reasons: with a appropriate ratio, the intensity of the phase reference pulse is not too strong or too weak, thus easier for modulation; the response frequency of the modulators are limited, we need to modulate with appropriate frequencies.

A total of $N_{total}$ signal pulses are sent for different fiber distances. The ratio of sending $X$ ($Z$) basis is $p_X$ ($p_Z$). In the $X$ basis, the ratio of sending the vacuum and decoy states, $0$, $\mu_1$ and $\mu_2$ are $p_0$, $p_1$ and $p_2$. In Z basis, the fractions of ``sending'' and ``not-sending'' pulses are $p_{z1}$ and $p_{z0}$, respectively. We used fixed ratios because of amplifier and modulator frequency response limitations. The cutoff frequencies of our amplifiers and modulators are between 75 kHz and 10 GHz, so their responses would become nonlinear if no signals are produced for a long time, as could happen if the fraction of $Z$ basis is high. To avoid this scenario, we sent a fixed  fraction of pulses in the $X$ basis. With better electronic components, we could optimize the experimental parameters to further enhance the final key rate.

Finally, based on the above parameters, we calculated the intensity at Alice's (Bob's) output. We also estimated the detection counts, taking the optics and detection efficiencies into account. However, for high detection counts, the detector dead time further decreased detection efficiency, and the peak phase reference pulse intensities is around $2\sim5$ photons per 100 ns, which is relatively high. Thus, the actual detection counts are smaller than the theoretical estimates. However, note that this effect does not affect the detection efficiency of the signal pulses.

\begin{table*}[htb]
\centering
  \caption{Experimental parameters for the different fiber lengths, the distances with a star marker stand for the second experimental test with the improved system.}
\begin{tabular}{c|cccc|ccc}
\hline
Fiber Length & 0 km & 50 km & 100 km & 150 km & 100 km$^*$ & 200 km$^*$ & 300 km$^*$\\
\hline
$\mu_1$		& 0.050	& 0.050	& 0.050	& 0.050 & 0.100 & 0.100 & 0.100 \\
$\mu_2$		& 0.159	& 0.160	& 0.177	& 0.197 & 0.200 & 0.200 & 0.200 \\
$\mu_z$		& 0.557	& 0.480	& 0.452	& 0.433 & 0.425 & 0.425 & 0.425 \\
$\mu_{ref}$	& 0.156	& 0.371	& 1.173	& 0.824 & 0.595 & 5.30  & 47.2  \\ 
\hline
$p_X$		& 0.30	& 0.30	& 0.30	& 0.30  & 0.20 & 0.20 & 0.20 \\
$p_Z$		& 0.70	& 0.70	& 0.70	& 0.70  & 0.80 & 0.80 & 0.80 \\
$p_0$		& 0.25	& 0.25	& 0.25	& 0.25  & 0.20 & 0.20 & 0.20 \\
$p_1$		& 0.50	& 0.50	& 0.50	& 0.50  & 0.60 & 0.60 & 0.60 \\
$p_2$		& 0.250	& 0.25	& 0.25	& 0.250 & 0.20 & 0.20 & 0.20 \\
$p_{z0}$	& 0.978	& 0.978	& 0.978	& 0.978 & 0.958 & 0.958 & 0.958 \\
$p_{z1}$	& 0.022	& 0.022	& 0.022	& 0.022 & 0.042 & 0.042 & 0.042 \\
\hline
Reference Width	(ns)			& 100	& 100	& 100	& 300   & 500   & 500   & 500    \\
Output Intensity (pW)		& 0.98  & 1.97  & 6.07  & 19.05 & 15.25  & 135.61 & 1208.27 \\ 
Target Detections (MHz)		& 4.36	& 3.11	& 3.03	& 3.01  & 4.00   & 4.00   & 4.00    \\
\hline
\end{tabular}
\label{Tab:Parameters}
\end{table*}

\section{Detailed Experimental Results}

We characterized our experimental system and then performed experimental tests with several different fiber lengths. The results are summarized in Tabs.~\ref{Tab:Characterization} and Tab.~\ref{Tab:Result}. First, we test the SNS-TF-QKD with a detector dark count probability of approximately $10^{-6}$ per pulse and  X-basis error rate of approximately 10\%. The experiment was performed with fiber distances of 0 km, 50 km, 100 km and 150 km.

Next, we improve the system using a SNSPD with a lower detector dark count probability of approximately $10^{-7}$ per pulse. We also used a linear voltage amplifier for the phase modulator and modify the phase calculation algorithm, so the phase estimation is better and the X-basis baseline error rate drops to approximately 2\%. Note that the source and the fiber conditions are not changed, so the phase drift rate remains the same as in the first test. The SNS-TF-QKD was tested under this scenario for the fiber distances of 100 km, 200 km and 300 km. For all the tables showing the experimental parameters and results, we mark the distance in this scenario with a star marker ``*'' for better distinguishability.

Tab.~\ref{Tab:Characterization} shows the results for the fiber's the transmittance and optical elements, as well as the  SNSPD efficiencies. For the optical elements, we measured the optical transmittance including the polarization controller (PC), the polarization beam splitter (PBS), and the beam splitter (BS). Here, transmittances are given for each of the two inputs (A/B) and outputs (ch1/ch2), as appropriate. The SNSPD efficiency measurements included the PC efficiency.

\begin{table*}[htb]
\centering
  \caption{Experimental parameters for different fiber lengths, the distances with a star marker stand for the second experimental test with the improved system.}
\begin{tabular}{c|cccc|ccc}
\hline
Fiber Length & 0 km & 50 km & 100 km & 150 km & 100 km$^*$ & 200 km$^*$ & 300 km$^*$ \\
$\eta_{FiberA}$	& 1	& 0.315 & 0.112 	 & 0.038  & 0.115  & 0.013  & 0.0015 \\
$\eta_{FiberB}$	& 1	& 0.316 & 0.120  & 0.033  & 0.103  & 0.013  & 0.0015 \\
\hline
PC-A & \multicolumn{7}{c}{94.2\%}\\
PC-B & \multicolumn{7}{c}{92.8\%}\\
\hline
PBS-A & \multicolumn{7}{c}{91.1\%}\\
PBS-B & \multicolumn{7}{c}{86.5\%}\\
\hline
BS-A-ch1 & \multicolumn{7}{c}{36.9\%}\\
BS-A-ch2 & \multicolumn{7}{c}{38.6\%}\\
BS-B-ch1 & \multicolumn{7}{c}{39.1\%}\\
BS-B-ch2 & \multicolumn{7}{c}{41.4 \%}\\
\hline
SNSPD-ch1 & \multicolumn{4}{c}{75.3\%} & \multicolumn{3}{c}{58.0\%}\\
SNSPD-ch2 & \multicolumn{4}{c}{76.6\%} & \multicolumn{3}{c}{38.0\%}\\
dark count & \multicolumn{4}{c}{$\sim 10^{-6}$} & \multicolumn{3}{c}{$\sim 10^{-7}$}\\
\hline
\end{tabular}
\label{Tab:Characterization}
\end{table*}

Tab.~\ref{Tab:Result} summarizes the experimental results. For each fiber length, the table shows the single photon yield $s_1$, phase error rate $e_1^{ph}$, and the final key rate $R$ for the best possible parameters, including the accepted phase difference $Ds$ (in degrees), and the parameter describing the phase estimation success probability $rc$. The fraction of the data accepted with the smaller $rc$ value is denoted by the parameter $r_{rc}$. In our experimental implementation, we applied a digital window to select the signal in the middle of each pulse, where the interference is expected to be better. The fraction of the data falling within this window is represented by the parameter $r_{gate}$. Note that all the detections are filtered according to the digital window and the estimation success probability $rc$ by Charlie, before the detections are announced.

This table also summarizes the raw data used for the calculations. The total number of signal pulses is given by $N_{total}$, and the error rates in the $Z$ and $X$ bases are given by ``QBER(Z)'' and ``QBER(X11)''/``QBER(X22)'', for the decoy states ``11'' and ``22''. The following rows list the numbers of pulses Alice and Bob sent in different decoy states, labelled as ``Sent-ABCD'', where ``A'' (``B'') is ``X'' or ``Z'' indicating the basis Alice (Bob) uses; ``C'' (``D'') is ``0'', ``1'',  ``2'' or  ``3'', indicating the intensity Alice (Bob) uses is ``vacuum'', ``$\mu_1$'', ``$\mu_2$'' or ``$\mu_z$''. The total number of pulses Alice and Bob sent is listed as ``Sent-AB''. As with the the numbers of pulses sent,  the numbers of detections are listed as ``Detected-ABCD''. The total valid detections reported by Charlie is denoted as ``Detected-Valid-ch'', where ``ch'' can be ``Det1'' or ``Det2'' indicating by which detector the counts are recorded. The table also gives the numbers of detections falling within the accepted difference range $Ds$, listed as ``Detected-ABCD-Ds-Ch'', where ``Ds'' indicates that only the data within the accepted range $Ds$ are counted, ``Ch'' indicates the detection channel. The numbers of correct detections are listed as ``Correct-ABCD-Ds-Ch'', and are used to calculate the $X$ basis error rate. The optimized acceptance ranges are listed on the top lines of this table.

\begin{table*}[htb]
\centering
  \caption{Experimental results for different fiber lengths, the distances with a star marker stand for the second experimental test with the improved system.}
\begin{tabular}{c|cccc}
\hline
Fiber Length & 0 km & 50 km & 100 km & 150 km \\
$R$			& $9.496\times 10^{-5}$ & $2.281\times 10^{-5}$ & $7.664\times 10^{-6}$ & $1.715\times 10^{-6}$ \\
$s_1$		& $2.20\times 10^{-1}$  & $7.23\times 10^{-2}$  & $2.57\times 10^{-2}$  & $7.45\times 10^{-3}$  \\
$e_1^{ph}$	& 13.3\% & 13.1\% & 15.0\% & 14.1\% \\
\hline
QBER(Z) 	& 2.11\% & 2.31\% & 2.58\% & 3.09\% \\
QBER(X11)	& 12.0\% & 11.7\% & 13.0\% & 12.0\% \\
QBER(X22)	& 12.0\% & 11.6\% & 12.9\% & 12.2\% \\
\hline
$rc$		& 0.05 			& 0.03 			& 0.05			& 0.04 			\\
$Ds$		& $10^\circ$		& $10^\circ$ 	& $10^\circ$ 	& $10^\circ$		\\
$r_{rc}$	& 0.41			& 0.32			& 0.43			& 0.42 			\\
$r_{gate}$	& 0.46 			& 0.46			& 0.47			& 0.44 			\\
\hline
$N_{total}$	& \multicolumn{4}{c}{$7.20\times 10^{11}$}\\
\hline
Sent-ZZ		& 352427400000	& 352525000000	& 352173800000	& 352447400000	\\
Sent-ZX00	& 37133800000	& 37066400000	& 37246200000	& 37102600000	\\
Sent-ZX01	& 73965600000	& 73930800000	& 73866200000	& 73917400000	\\
Sent-ZX02	& 36939600000	& 36916200000	& 37092800000	& 36877600000	\\
Sent-ZX30	& 860000000		& 851600000		& 857800000		& 855600000		\\
Sent-XZ00	& 36989000000	& 36995200000	& 37046000000	& 37074400000	\\
Sent-XZ10	& 73861400000	& 73866400000	& 73981400000	& 73915200000	\\
Sent-XZ20	& 36925200000	& 36958600000	& 36986000000	& 36998800000	\\
Sent-XZ03	& 826800000		& 833000000		& 864200000		& 838800000		\\
Sent-XX00	& 4034200000		& 4059800000		& 4006800000		& 4003000000		\\
Sent-XX01	& 8159800000		& 8127400000		& 8009600000		& 8139000000		\\
Sent-XX02	& 4088200000		& 4040600000		& 4015000000		& 4087000000		\\
Sent-XX10	& 8140200000		& 8120200000		& 8084400000		& 8135600000		\\
Sent-XX20	& 4088800000		& 4101600000		& 4083200000		& 4070400000		\\
Sent-XX11	& 16166800000	& 16162600000	& 16152400000	& 16118600000	\\
Sent-XX22	& 4038600000		& 4060400000		& 4071400000		& 4064400000		\\
\hline
\end{tabular}
\label{Tab:Result}
\end{table*}
\addtocounter{table}{-1}

\begin{table*}[htb]
\centering
  \caption{Experimental results for different fiber lengths, the distances with a star marker stand for the second experimental test with the improved system.}
\begin{tabular}{c|cccc}
\hline
Fiber Length 		& 0 km & 50 km & 100 km & 150 km \\
Detected-Valid-Det1	& 3180640457		& 1114550015		& 367608430	& 121134943  \\	
Detected-Valid-Det2	& 3351567620		& 1188135091		& 395452379	& 131162147 \\
\hline
Detected-ZX00	& 33179		& 18044		& 23574		& 18883			 \\
Detected-ZX01	& 253374727	& 74427515	& 31550408	& 9356926		 \\
Detected-ZX02	& 402438421	& 114186348	& 55219631	& 18763685		 \\
Detected-ZX30	& 30774678 	& 7654434	& 3160510	& 983736			 \\
Detected-XZ00	& 33316		& 18000		& 23153		& 18781			 \\
Detected-XZ10	& 243332100	& 70866220	& 28911062	& 9001415		 \\
Detected-XZ20	& 383172746	& 110124270	& 52860174	& 18303719		 \\
Detected-XZ03	& 30611835 	& 7574806	& 3185989	& 923116			 \\
Detected-XX00	& 3563		& 1910		& 2529		& 2047			 \\
Detected-XX01	& 27973801	& 8192443	& 3419006	& 1028370		 \\
Detected-XX02	& 44545111	& 12516886	& 5981633	& 2084672		 \\
Detected-XX10	& 26835876	& 7786310	& 3154146	& 990573			 \\
Detected-XX20	& 42404015	& 12207818	& 5845716	& 2004057		 \\
Detected-XX11	& 107700658	& 31698647	& 13173657	& 3992955		 \\
Detected-XX22	& 83470873	& 24461170	& 11799518	& 4042977		 \\
\hline
Detected-XX11-Ds-Ch1	& 6030273	& 1782747	& 738886		& 222421		 \\
Detected-XX11-Ds-Ch2	& 6476445	& 1909673	& 797997		& 241777		 \\
Correct-XX11-Ds-Ch1		& 5302155	& 1574536	& 642684		& 195446		 \\
Correct-XX11-Ds-Ch2		& 5702371	& 1687713	& 694483		& 212980		 \\
\hline
Detected-ZZError	& 12154186	& 3297796	& 1512793	& 556383		\\
Detected-ZZCorrect	& 563284450	& 139662278	& 57064249	& 17435775	\\
\hline
\end{tabular}
\label{Tab:Result2}
\end{table*}
\addtocounter{table}{-1}

\begin{table*}[htb]
\centering
  \caption{Experimental results for different fiber lengths, the distances with a star marker stand for the second experimental test with the improved system.}
\begin{tabular}{c|ccc}
\hline
Fiber Length & 100 km$^*$ &	200 km$^*$ & 300 km$^*$\\
$R$			& $1.841\times10^{-4}$ & $2.405\times 10^{-5}$ & $1.957\times 10^{-6}$\\
$s_1$		& $3.26\times10^{-2}$  & $3.96\times10^{-3}$  & $4.99\times10^{-4}$ \\
$e_1^{ph}$	& 2.09\% & 2.07\% & 3.58\% \\
\hline
QBER(Z) 	&  4.15\% & 4.25\% & 5.29\% \\
QBER(X11)	&  1.49\% & 1.43\% & 2.43\% \\
QBER(X22)	&  1.44\% & 1.33\% & 2.45\% \\
\hline
$rc$		& 1				& 1				 & 0.3 		\\
$Ds$		& $5^\circ$		& $5^\circ$		& $8^\circ$ \\
$r_{rc}$	& 1				& 1				 & 0.99985 		\\
$r_{gate}$	& 0.87			& 0.85			 & 0.82		\\
\hline
$N_{total}$	& \multicolumn{3}{c}{$7.20\times 10^{11}$}\\
\hline
Sent-ZZ		& 460668400000	& 460820400000	& 460688800000	\\
Sent-ZX00	& 22049000000	& 422964400000	& 22106000000	\\
Sent-ZX01	& 66513800000	& 66391000000	 & 66484600000	\\
Sent-ZX02	& 22114400000	& 22093200000	 & 22077000000	\\
Sent-ZX30	& 979600000		& 977400000		 & 975200000		\\
Sent-XZ00	& 22110600000	& 22134600000	 & 22116200000	\\
Sent-XZ10	& 66190600000	& 66223400000	 & 66266600000	\\
Sent-XZ20	& 22113000000	& 22121600000	 & 22149800000	\\
Sent-XZ03	& 967400000		& 959800000 	 	& 961200000		\\
Sent-XX00	& 1139200000 	& 1147000000 	& 1155000000	\\
Sent-XX01	& 3463800000		& 3481800000 	& 3460800000	\\
Sent-XX02	& 1148200000		& 1161000000 	& 1163600000	\\
Sent-XX10	& 3450000000		& 3456000000 	& 3454200000	\\
Sent-XX20	& 1162800000		& 1155400000 	& 1157600000	\\
Sent-XX11	& 10386200000	& 10391400000 	 & 10378600000	\\
Sent-XX22	& 1137600000		& 1138600000 	& 1131600000	\\
\hline
\end{tabular}
\label{Tab:Result3}
\end{table*}
\addtocounter{table}{-1}

\begin{table*}[htb]
\centering
  \caption{Experimental results for different fiber lengths, the distances with a star marker stand for the second experimental test with the improved system.}
\begin{tabular}{c|ccc}
\hline
Fiber Length & 100 km$^*$ &	200 km$^*$ & 300 km$^*$\\
Detected-Valid-Det1	& 1002557125		& 116230879 & 14820318 \\	
Detected-Valid-Det2	& 658513000		& 74899337 	& 9447864 \\
\hline
Detected-ZX00	& 10303		& 4852		& 5020		 \\
Detected-ZX01	& 221965494	& 26391605	& 3349153	 \\
Detected-ZX02	& 146107910	& 17075224	& 2209525	 \\
Detected-ZX30	& 14479458	& 1602369	& 193636	 \\
Detected-XZ00	& 10417		& 5023		& 5002		 \\
Detected-XZ10	& 218715443	& 25863880	& 3351934	 \\
Detected-XZ20	& 149096872	& 17224771	& 2194551	 \\
Detected-XZ03	& 13794658	& 1529027	& 194968	 \\
Detected-XX00	& 526		& 253		& 250		 \\
Detected-XX01	& 11558494	& 1386715	& 174753	 \\
Detected-XX02	& 7594839	& 895573	& 115537	 \\
Detected-XX10	& 11403871	& 1351285	& 175531	 \\
Detected-XX20	& 7836356	& 901455	& 114681	 \\
Detected-XX11	& 68766956	& 8164375	& 1040779	 \\
Detected-XX22	& 15102659	& 1760343	& 224088	 \\
\hline
Detected-XX11-Ds-Ch1	& 2534611	& 303772	 & 59786 \\
Detected-XX11-Ds-Ch2	& 1660965	& 195771	 & 38201 \\
Correct-XX11-Ds-Ch1		& 2497028	& 299434	 & 58301 \\
Correct-XX11-Ds-Ch2		& 1635909	& 192962	 & 37301 \\
\hline
Detected-ZZError	& 23272506	& 2660102	& 415203  \\
Detected-ZZCorrect	& 537963242	& 59913256	& 7426487 \\
\hline
\end{tabular}
\label{Tab:Result4}
\end{table*}

Different phase difference ranges $Ds$ and accepted phase estimation success probabilities $rc$ yield different $X$ basis QBERs and detection counts. For Tab.~\ref{Tab:Result}, we optimized these values to a maximize the final key rate. Taking the 150 km data as an example, the $X$ basis QBERs when Alice and Bob sent decoy states $\mu_1$ and $\mu_2$ are listed in Tabs.~\ref{Tab:ResultQBERXX11} and \ref{Tab:ResultQBERXX22}, respectively, and the numbers of detections in Tabs.~\ref{Tab:ResultDetXX11} and \ref{Tab:ResultDetXX22}. Finally, Tab.~\ref{Tab:ResultKeyRate} summarize the secure key rates achieved for different parameter values. these key rates were calculated by searching though ranges of parameter values to find the optimized key rate. For this data set, we obtained the optimized key rate with $Ds/2$=$10^\circ$ and $rc=0.04$, where $Ds/2$ is half the phase difference range.

We need to point out the post-selection that based on the phase estimation success probabilities $rc$ is necessary for extracting valid results when the phase drift cannot be fully compensated. But this might not be needed if the phase estimation is good enough. In our second experimental trial with the lower detector dark count probability and smaller X-basis baseline error rate, the X-basis error rate are almost the same for $rc=0.01$ and $rc=1$, which is around 2\%. Thus we can set $rc=1$ without dropping any detections due to inaccurate phase compensation. In our analysis, we optimize the final key rate using the best parameters.

\begin{table*}[htb]
\centering
  \caption{$X$ basis QBERs for decoy state $\mu_1$ with 150 km of fiber.}
\begin{tabular}{ccccccccc}
\hline
RC$\mid$Ds/2 & deg=1$^\circ$	& deg=2$^\circ$	& deg=4$^\circ$	& deg=6$^\circ$	& deg=8$^\circ$	& deg=10$^\circ$	& deg=12$^\circ$	& deg=15$^\circ$\\
\hline
rc=0.01	& 9.60\% & 9.70\% & 9.70\% & 9.80\% & 9.90\% & 9.90\% & 10.0\% & 10.2\% \\	
rc=0.02	& 10.6\% & 10.6\% & 10.6\% & 10.6\% & 10.7\% & 10.8\% & 10.8	\% & 11.0\% \\
rc=0.04	& 11.9\% & 12.0\% & 11.9\% & 11.9\% & 12.0\% & 12.0\% & 12.1	\% & 12.2\% \\	
rc=0.05	& 12.5\% & 12.6\% & 12.5\% & 12.5\% & 12.5\% & 12.6\% & 12.6	\% & 12.8\% \\
rc=0.10	& 14.9\% & 15.0\% & 14.8\% & 14.8\% & 14.8\% & 14.8\% & 14.9	\% & 15.0\% \\
\hline
\end{tabular}
\label{Tab:ResultQBERXX11}
\end{table*}

\begin{table*}[htb]
\centering
  \caption{$X$ basis QBERs for decoy state $\mu_2$ with 150 km of fiber.}
\begin{tabular}{ccccccccc}
\hline
RC$\mid$Ds/2 & deg=1$^\circ$	& deg=2$^\circ$	& deg=4$^\circ$	& deg=6$^\circ$	& deg=8$^\circ$	& deg=10$^\circ$	& deg=12$^\circ$	& deg=15$^\circ$\\
\hline
rc=0.01	& 9.60\% & 9.60\% & 9.80\% & 9.90\% & 10.0\% & 10.0\% & 10.2\% & 10.4\% \\
rc=0.02	& 10.5\% & 10.5\% & 10.5\% & 10.6\% & 10.7\% & 10.8\% & 10.9\% & 11.1\% \\
rc=0.04	& 12.0\% & 12.0\% & 12.0\% & 12.1\% & 12.2\% & 12.2\% & 12.2\% & 12.4\% \\
rc=0.05	& 12.6\% & 12.6\% & 12.6\% & 12.6\% & 12.7\% & 12.7\% & 12.8\% & 12.9\% \\
rc=0.10	& 14.8\% & 14.9\% & 14.9\% & 14.9\% & 15.0\% & 14.9\% & 15.0\% & 15.1\% \\
\hline
\end{tabular}
\label{Tab:ResultQBERXX22}
\end{table*}

\begin{table*}[htb]
\centering
  \caption{$X$ basis Detections for decoy state $\mu_1$ with 150 km of fiber.}
\begin{tabular}{ccccccccc}
\hline
RC$\mid$Ds/2 & deg=1$^\circ$	& deg=2$^\circ$	& deg=4$^\circ$	& deg=6$^\circ$	& deg=8$^\circ$	& deg=10$^\circ$	& deg=12$^\circ$	& deg=15$^\circ$\\
\hline
rc=0.01	& 25920 & 43810  & 77901  & 112095 & 146611 & 181930 & 216085 & 269071  \\
rc=0.02	& 42331	& 71760  & 129987 & 188283 & 248574 & 309691 & 370416 & 462085  \\
rc=0.04	& 63617 & 107070 & 193671 & 282255 & 372405 & 464198 & 555951 & 693152  \\
rc=0.05 & 70817 & 119687 & 216583 & 314920 & 415225 & 517508 & 619325 & 772151  \\
rc=0.10 & 94310 & 159293 & 287220 & 417054 & 549170 & 682478 & 815517 & 1015767 \\
\hline
\end{tabular}
\label{Tab:ResultDetXX11}
\end{table*}

\begin{table*}[htb]
\centering
  \caption{$X$ basis Detections for decoy state $\mu_2$ with 150 km of fiber.}
\begin{tabular}{ccccccccc}
\hline
RC$\mid$Ds/2 & deg=1$^\circ$	& deg=2$^\circ$	& deg=4$^\circ$	& deg=6$^\circ$	& deg=8$^\circ$	& deg=10$^\circ$	& deg=12$^\circ$	& deg=15$^\circ$	\\
\hline
rc=0.01 & 25623 & 43760  & 77889  & 112020 & 146619 & 181508 & 215655 & 268197  \\
rc=0.02 & 41921 & 71575  & 129710 & 188161 & 248720 & 309052 & 369819 & 460779  \\	
rc=0.04 & 63038 & 106851 & 193420 & 281872 & 372448 & 463651 & 555692 & 692316  \\	
rc=0.05 & 70180 & 119398 & 216221 & 314377 & 415230 & 516792 & 618938 & 771397  \\	
rc=0.10 & 93939 & 159293 & 287262 & 417313 & 550154 & 682961 & 816726 & 1016446 \\	
\hline
\end{tabular}
\label{Tab:ResultDetXX22}
\end{table*}

\begin{table*}[htb]
\centering
  \caption{Key Rate for different parameters with 150 km fiber.}
\begin{tabular}{ccccccccc}
\hline
RC$\mid$Ds/2 & deg=1$^\circ$	& deg=2$^\circ$	& deg=4$^\circ$	& deg=6$^\circ$	& deg=8$^\circ$	& deg=10$^\circ$	& deg=12$^\circ$	& deg=15$^\circ$\\
\hline
rc=0.01	& $0$ & $4.71\times 10^{-7}$ & $8.29\times 10^{-7}$ & $9.51\times 10^{-7}$ & $9.97\times 10^{-7}$ & $1.02\times 10^{-6}$ & $1.03\times 10^{-6}$ & $1.01\times 10^{-6}$ \\
rc=0.02	& $0$ & $7.62\times 10^{-7}$ & $1.28\times 10^{-6}$ & $1.47\times 10^{-6}$ & $1.51\times 10^{-6}$ & $1.52\times 10^{-6}$ & $1.52\times 10^{-6}$ & $1.49\times 10^{-6}$ \\
rc=0.04	& $0$ & $5.84\times 10^{-7}$ & $1.39\times 10^{-6}$ & $1.61\times 10^{-6}$ & $1.70\times 10^{-6}$ & $1.71\times 10^{-6}$ & $1.71\times 10^{-6}$ & $1.69\times 10^{-6}$ \\
rc=0.05	& $0$ & $3.10\times 10^{-7}$ & $1.23\times 10^{-6}$ & $1.49\times 10^{-6}$ & $1.60\times 10^{-6}$ & $1.61\times 10^{-6}$ & $1.62\times 10^{-6}$ & $1.58\times 10^{-6}$ \\
rc=0.10 & $0$ & $0$ & $1.17\times 10^{-8}$ & $4.05\times 10^{-7}$ & $5.71\times 10^{-7}$ & $6.40\times 10^{-7}$ & $6.67\times 10^{-7}$ & $6.44\times 10^{-7}$ \\\hline
\end{tabular}
\label{Tab:ResultKeyRate}
\end{table*}

\bibliography{BibSNSTFQKD}

\end{document}